\documentclass[10pt,twocolumn]{article}
\usepackage{graphicx}
\usepackage{amsmath,amsthm,amssymb}
\usepackage{ccaption}
\usepackage{bigdelim,multirow}

\newtheorem{ex}{Example}
\newtheorem{step}{Step}

\title{Stochastic approach to molecular interactions and 
computational theory of metabolic and genetic regulations}
\author{H. Kimura, H. Okano and R.J. Tanaka\footnote{Corresponding author.
Tel.: +81(52)736-5861, Fax: +81(52)736-5862, Email: reiko@bmc.riken.jp}}
\date{Bio-Mimetic Control Research Center, RIKEN,\\
Shimo-shidami, Moriyama-ku, Nagoya 463-0003, Japan}

\begin{document}
\maketitle

\begin{abstract}
Binding and unbinding of ligands to specific sites of a macromolecule are
one of the most elementary molecular interactions inside the cell that embody 
the computational processes of biological regulations.
The interaction between transcription factors and the operators of genes
and that between ligands and binding sites of allosteric enzymes are
typical examples of such molecular interactions.
In order to obtain the general mathematical framework of biological
regulations, we formulate these interactions as finite Markov processes
and establish a computational theory of regulatory activities of macromolecules
based mainly on graphical analysis of their state transition diagrams.
The contribution is summarized as follows:

\noindent
(1) Stochastic interpretation of Michaelis-Menten equation is given. \\
(2) Notion of \textit{probability flow} is introduced in relation to 
detailed balance.\\
(3) A stochastic analogy of \textit{Wegscheider condition} is given 
in relation to loops in the state transition diagram. \\
(4) A simple graphical method of computing the regulatory activity in terms of
ligands' concentrations is obtained for Wegscheider cases. \\
(5) A general comutational scheme of the stationary probability distribution
is obtained in terms of probability flows.
\end{abstract}

\noindent
\textit{keywords}: Biological regulation, molecular interaction, Markov process, 
stationary probability distribution, probability flow

\section{Introduction}

Control is a ubiquitous built-in mechanism supporting a variety of functions of living organisms. The advent of cybernetics in the late 1940's established a close link between control in living organisms and that of man-made artifact. Notions like feedback, feedforward, robustness, stability, and noise, which are fundamental in control engineering turned out to be relevant to the control mechanisms of living organisms.  Rapid progress in molecular biology in the last fifty years opened up a new world of intracellular control of biochemical processes including those of metabolism and gene expression. Studying control mechanisms became an important issue of biology. Monod correctly called this field \textit{microscopic cybernetics} \cite{monod}, while others called it \textit{regulatory biology} \cite{yanofsky}. The continuing endeavor to clarify the material basis of regulations reveals vast complexity of regulatory networks associated with huge variety of material links inside a cell, which in turn motivated considerable interest in the quantitative analysis of regulation mechanisms through mathematical modeling and simulation. 
The modeling/simulation paradigm has been used to reveal the underlying structural properties of complex regulatory networks.  
Existence of cellular switches \cite{ackers}\cite{cherry}\cite{gardner}\cite{killer}\cite{mochizuki}\cite{ozbudak}\cite{ptashne}, evaluation of molecular fluctuations \cite{gilles}\cite{kaern}\cite{mcadams}\cite{ordenaa}, analysis of biological oscilations \cite{bliss}\cite{chen}\cite{griffith}, consideration of network robustness \cite{kitano}\cite{yi}, theoretical demonstration of cellular behaviors \cite{endy}\cite{mca}\cite{savageau}, are examples of its contributions. Excellent reviews of these achievements are found in \cite{smolen}\cite{thomas}\cite{tyson}.
Recently, we proposed the notion of compound control \cite{tanaka} which captures a salient characteristic feature of biological control. Compound control is a biological way of realizing proper input/output relationship of regulatory dynamics to cope with diverse, complicated and unpredictable environmental changes. 

It is paradoxical that the theoretical and/or physical basis of the modeling/simulation paradigm has not been well exploited, in spite of the remarkable success it has attained.  For instance, in many papers on intracellular regulations, the traditional Michaelis-Menten-Hill equation in enzymology is extensively used for describing molecular interactions, e.g., the action of repressors in transcription regulation, or the effect of feedback \cite{bliss}\cite{gardner}\cite{mackey} \cite{mochizuki}\cite{ozbudak}\cite{wong}, without sound justification.  

The transcription rate is regulated through binding of transcription 
factors in some domains of DNA,
the so-called \textit{cis}-regulatory elements \cite{bolouri}, that are able to affect  transcription initiation. The affinity of \textit{cis}-regulatory elements to the transcription factors determines the overall transcription rate of the gene. They are correlated with each other and collectively exhibit synergistic effects through which a complex computation is performed to achieve adequate transcription control \cite{lloyd}. 
These factors are represented in the context of a thermodynamical equilibrium energy distribution suggesting to plausible models of transcription regulation \cite{berg}\cite{bintu}\cite{panh}\cite{wolfd}.
This paper aims to establish a unified framework with solid mathematical grounding and physical plausibility that is able to capture various bio-physical attributes of intracellular regulations. 

A significant portion of genes encode enzymes which are involved in metabolic control, another important form of intracellular regulation. Metabolic control is a relatively old subject of bio-chemistry, where allosteric enzymes are the major agents of control. Allosteric enzyme is a macro-molecule that has several binding sites which can bind substrate and ligands. The binding pattern of ligands changes the conformations of the enzyme, through which the activity of the enzyme for the target reaction is controlled. Quantification of the activity changes of allosteric enzyme has been a central issue in biochemistry since the classical Monod-Wyman-Changeux model was first proposed \cite{monodj}\cite{wyman}. Since then, various theoretical and experimental methods have been used to predict the conformation changes due to substrate binding and modification by ligands \cite{berg}\cite{coppey}\cite{freire}\cite{gibson}\cite{gunasekaran}\cite{king}\cite{ming}\cite{han}\cite{segel}\cite{kampen}\cite{Zaslaver}.

The regulatory mechanism of allosteric enzymes is in some sense similar to that of operons, in spite of the obvious functional and structural differences between them. Both of them are controlled by interactions between binding sites (catalytic sites for metabolic cases and \textit{cis}-regulatory elements for genetic cases) and binding factors (ligands and substrate in the metabolic case and transcription factors in the genetic case). In the metabolic case, interactions take place between proteins, while in the genetic case they take place between protein and DNA. It is interesting that although the allosteric protein is a player in both regulations, it plays totally different roles. In the metabolic control, it is controlled by ligands, while in the genetic control it acts as transcription factors to control genes.

The protein/protein interactions of metabolic regulation and protein/DNA interactions in transcription control are weak and essentially reversible, 
and they share many properties.
In fact, equilibrium statistical mechanics is used to quantify both
metabolic regulation \cite{gibson} and genetic regulation \cite{bintu}\cite{panh}\cite{wolfd}.
If we view the binding/unbinding processes taking place between binding sites and binding factors in both cases as a common stochastic process, it suggests a common theoretical framework to deal with metabolic and genetic regulations in a unified way. 

Along this line, we propose a common theoretical framework for describing biological regulators of metabolic and genetic regulations 
to facillitate quantitative descriptions of intracellular bio-chemical processes.
The core of our approach is to describe protein/protein and protein/DNA interactions with a common Markov process focusing on the behavior of a single molecule rather than a population of molecules. This viewpoint enables us to describe phenomena with a finite state Markov process, which greatly simplifies 
the argument by avoiding the difficulties associated with infinite dimensionality
\cite{munsky}.
All the quantification features associated with various complicated physical and biological phenomena
can then be condensed into transition probabilities.
We introduce a new notion of probability flow that offers a new insight
into the stationary distribution of the Markov process.
The stationary equation is regarded as representing the
conservation of probability flows at each node.
It is closely related to the Wegscheider condition derived a century ago.
Based on the notion of probability flow, we derive a representation of
the stationary probability distribution, which leads to a unified quantitative form
for the general biological regulator.
Our method of computing the stationary probability distribution
dramatically simplifies the classical King and Altman method \cite{king},
which is used for computing the stationary distribution of chemical processes. 
It is expected to yield a new computational tool for operon regulation
that can deal with the increasing complexity of operons \cite{santillan}.

Metabolic and genetic controls have been investigated relatively independently 
because of their apparent large differences in modality. 
Therefore, there have been few studies dealing with systems that combine metabolic  and genetic regulations \cite{Zaslaver}. The unified framework described in this paper will enhance our understanding of intracellular biochemical processes. 

In the next section, we formulate a biological regulator in an abstract way as a Markov process.  The abstract biological regulator is defined through the stationary solution of the Master Equation.  In Section \ref{sec3}, we introduce the notion of probability flow and explain its relevance to the stationary distribution. 
Sections \ref{sec4} and \ref{sec5} deal with the single loop case and introduce the Wegscheider condition, which guarantees that the probability flow vanishes.  Section \ref{sec6} generalizes the results of the preceding section to multi-loop cases.  
Section \ref{sec7} discusses the meaning of transition probabilities to deduce
biologically meaningful representations of biological regulations.

\section{Characterization of biological regulations as a
finite-state Markov process}

The main stage of genetic control is transcription regulation
whose fundamental mechanism is to change the transcription rate
of the operon via binding/unbinding of transcription factors
to and from their \textit{cis}-regulatory elements.
The main agent of metabolic control is the allosteric enzyme
that determine the rate of the target chemical reaction
via binding/unbinding of substrate and ligands.
The fundamental common feature of both regulations lies in
the molecular interaction between sites of the macro-molecule and
molecular binding factors that work through binding/unbinding processes
(see Table~\ref{table:1}).
      
\begin{table*}[bt]
\caption{
Comparison of genetic and metabolic regulations.}\label{table:1}
\begin{center}
    \begin{tabular}{c|p{2.6cm}|p{2.6cm}|p{2.6cm}|p{2.6cm}}  
  & \raisebox{-0.5\normalbaselineskip}{Agent} & 
    \raisebox{-0.5\normalbaselineskip}{Site} & 
    \raisebox{-0.5\normalbaselineskip}{Factor} &  Control\par Parameters  \\
    \hline \raisebox{-0.5\normalbaselineskip}{Genetic} & 
    \raisebox{-0.5\normalbaselineskip}{Operons} &  
    \textit{Cis}-regulatory elements &  Transcription factors & Transcription rate \\
    \hline \raisebox{-0.5\normalbaselineskip}{Metabolic} &  
    Allosteric \par enzymes  & \raisebox{-0.5\normalbaselineskip}{Catalytic sites} & 
    Ligands and \par substrates  & \raisebox{-0.5\normalbaselineskip}{Reaction rate}
      \end{tabular}
\end{center}
\end{table*}


These processes are obviously random subject to certain thermodynamical
constraints.
Our idea is to establish a mathematical framework to describe
such regulatory mechanisms in a unified way.
We now formulate the biological regulation in a somewhat
abstract way. We assume that the macro-molecule, which is
the main agent of the regulation, has $n$ binding sites (BS),
each of which can bind some of $m$ ligands with different 
affinities that depend on the binding patterns of sites.
Instead of the ligand, we use the word \textit{binding factors} (BF)
to emphasize the bilateral symmetry of sites to be bound
and factors to bind.
Each BS can bind only one BF at each time.
The binding sites are denoted by $b_{1},b_{2},\cdots,b_{n}$, 
while the binding factors by $U_{1},U_{2},\cdots,U_{m}$.
The state of the regulator is denoted by $n$-tuples of $m$
alphabet $U_{1},U_{2},\cdots,U_{m}$ denoting the BFs being bound 
and $\phi $ which represents the empty (unoccupied) site. 
If the regulator has three BSs, for example,
$S=(U_{1}\phi U_{2})$ denotes the state with
$b_{1}$ and $b_{3}$ being bound by $U_{1}$ and $U_{2}$,
respectively, and with $b_{2}$ empty.
Let $N$ be the total number of non-empty states,
which is obviously finite.
Since we have the empty state $S_{0}=(\phi\phi\cdots\phi)$,
the total number of states is $N+1$.

The state of the regulator changes as its binding pattern
changes, i.e., as BFs bind to and dissociate from BSs.
These are clearly stochastic phenomena and are legitimately
described as stochastic process.
Let the transition probability from the state $S_{j}$ to $S_{i}$
during an infinitesimal time duration $\Delta t$ be denoted by
$q_{ij}\Delta t$.
All the quantitative features of the complex biochemical 
process associated with protein/protein interactions and
protein/DNA interactions including origomerization \cite{han}\cite{wyman},
conformational change \cite{freire}\cite{panh} and
target localization \cite{coppey} can be adequately
represented in terms of $q_{ij}$.       
For the sake of clarity, we assume that binding or 
unbinding of only one BF to and from a BS occurs during
the infinitesimal time duration $\Delta t$.
Thus, the transition from $(U_{1}\phi U_{2})$ to
$(U_{1}\phi U_{3})$ is regarded as consequtive 
transitions from $(U_{1}\phi U_{2})$ to $(U_{1}\phi \phi)$
and from $(U_{1}\phi \phi)$ to $(U_{1}\phi U_{3})$.
The transition probability for unit time $q_{ij}$ represents the rate
of reaction from $S_{j}$ to $S_{i}$, and sometimes
it is regarded as identical to the rate constant
of the corresponding reaction.
We define ${\bf A}_{i}$ to be the set of states that are accessible
from $S_{i}$ during the infinitesimal time duration $\Delta t$.
We call it the \textit{adjacent set} of $S_{i}$.

Let $p_{i}(t)$ be the probability that the regulator is 
in a  state $S_{i}$ at time $t$.
Then, the stochastic time-evolution of the regulator is
described by the \textit{chemical master equation} (CME),
\begin{equation}
\frac{dp_{i}}{dt}=\sum_{j\in \bf{A}_{i}}q_{ij}p_{j}-\sum_{j\in \bf{A}_{i}}q_{ji}p_{i}.
\quad i=0,1,\cdots,N.
\label{eq2-1}
\end{equation}
The notation $\sum_{j\in {\bf A}_{i}}$ denotes the sum of all $j$ 
such that $S_{j}\in {\bf A}_{i}$.
The first term of the right-hand side of (\ref{eq2-1})
represents the net probability coming from adjacent states to $S_{i}$ 
while the second term the net probability
leaving $S_{i}$ to adjacent states.
The CME (\ref{eq2-1}) is written in the matrix form as
\begin{equation}
\frac{dp(t)}{dt}=Qp(t)
\label{eq2-2}
\end{equation}
where $p(t)$ denotes the $(N+1)$-dimensional vector whose $(i+1)$-th
component is $p_{i}(t)$.
The matrix $Q$ in  (\ref{eq2-2}) is called a 
\textit{transition matrix}.

\begin{figure}[tb]
  \begin{center}
    \includegraphics[width=0.4\textwidth]{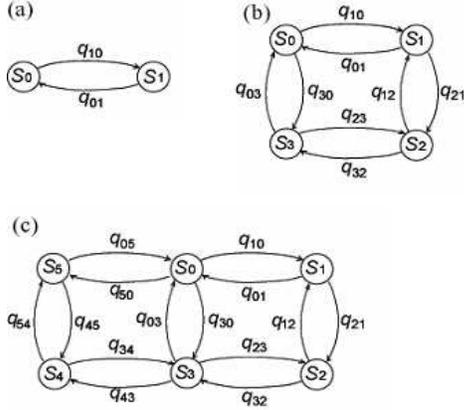}
  \end{center}
  \caption{State transition diagram of
(a) Example 1, (b) Example 2,(c) Example 3.}
  \label{fig:1}
\end{figure}

A finite Markov process can be described by the state-transition
diagram.
In Figure~\ref{fig:1} shows examples of state-transition diagrams.
The most salient feature of the Markov process for our purpose  
is the reflection property that $q_{ij}\neq 0$
always implies $q_{ji}\neq 0$ because of the reversibility of the molecular
interactions we are interested in this paper.
In terms of the adjacent set ${\bf A}_{i}$,
$S_{j}\in {\bf A}_{i}$ always implies $S_{i}\in{\bf A}_{j}$.

It is known \cite{kampen} that the solution of (\ref{eq2-2})
converges to a unique equilibrium stationary solution $p$
that satisfies
\begin{equation}
Qp=0,
\label{eq2-3}
\end{equation}
or equivalently,
\begin{equation}
\sum_{j\in {\bf A}_{i}}q_{ij}p_{j}-\sum_{j\in {\bf A}_{i}}q_{ji}p_{i}=0.
\label{eq2-4a}
\end{equation}
Equaiton (\ref{eq2-3}) or (\ref{eq2-4a}) is called
the \textit{stationary state equation} (SSE).
Usually, the convergence is faster than the chemical reactions 
controlled by the regulator.
Therefore, we can assume that the regulator is always in a
stationary distribution satisfying  (\ref{eq2-3}),
together with the normalization constraint,
\begin{equation}
\sum^{N}_{i=0}p_{i}=1.
\label{eq2-4}
\end{equation}

Let $\gamma_{i}$ be the activity of the regulator at state $S_{i}$.
The overall activity $\gamma$ of the regulator is then
defined as the average activity with respect to the 
stationary distribution $p$, i.e.,
\begin{equation}
\gamma=\sum^{N}_{i=0}\gamma_{i}p_{i}.
\label{eq2-5}
\end{equation}
The examples that follow show that the above definition of the regulation
activity adequately describes both genetic and metabolic regulations.

\begin{ex}\upshape
(Stochastic version of Michaelis-Menten equation)\\
Consider a molecule of an enzyme $E$ that catalyses the reaction
of producing a product $P$ from a substrate $S$, i.e.,
\begin{equation}
E+S \overset{k_{1}}{\underset{k_{-1}}{\rightleftarrows}} ES 
\overset{k_{2}}{\rightarrow} E+P.
\label{eq2-6}
\end{equation}
An enzyme molecule can be regarded as a biological regulator
with one BS and one BF (the substrate).
The enzyme has the two state $S_{0}$ and $S_{1}$,
the empty state and the state occupied by a substrate molecule,
respectively.
The transition diagram is shown in Fig.~\ref{fig:1}(a).
The transition matrix $Q$ in this case is written as
\begin{equation}
Q=\begin{bmatrix}-q_{10}&q_{01}\\
q_{10}&-q_{01}\end{bmatrix},
\label{eq2-7}
\end{equation}
and the stationary distribution is given by
\begin{equation}
p_{0}=\frac{1}{1+\frac{q_{10}}{q_{01}}},\quad p_{1}=\frac{\frac{q_{10}}{q_{01}}}{1+\frac{q_{10}}{q_{01}}}.
\label{eq2-8}
\end{equation}
Since $q_{01}$ represents the probability of unbinding the substrate $S$ from
the BS during unit time, we can identy it with $k_{-1}$.
In addition, the probability of binding $S$ to the BS is proportional
to the concentration $\left[S \right]$ of the substrate with $k_{1}$
being its rate coefficient. Thus,
\begin{equation}
q_{01}=k_{-1},\quad q_{10}=k_{1}\left[S \right],
\label{eq2-9}
\end{equation}
which gives
\begin{equation}
p_{0}=\frac{K_{1}}{K_{1}+\left[S \right]}, \quad
p_{1}=\frac{\left[S \right]}{K_{1}+\left[S \right]}, \quad
K_{1}=\frac{k_{-1}}{k_{1}}.
\label{eq2-10}
\end{equation}
Since $p_{1}$ denotes the probability of the state of the enzyme with $S$
being bound (ES), 
the total number of ES molecules is given by $p_{1}\left[E \right]_{T}$,
where $\left[E \right]_{T}$ denotes the concentration of the total enzyme, 
which is assumed to be constant.
Therefore, since the maximum reaction rate is given by
$k_{2}\left[E \right]_{T}$, the ratio of the reaction rate $\upsilon$ to its
maximum $\upsilon_{\max}$ is given by
\begin{equation}
\frac{\upsilon }{\upsilon_{\max}}=\frac{k_{2}p_{1}\left[E \right]_{T}}{k_{2}\left[E \right]_{T}}=p_{1}=\frac{\left[S \right]}{K_{1}+\left[S \right]},
\label{eq2-11}
\end{equation}
which is the celebrated \textit{Michaelis-Menten equation}.
Here, the rapid equilibrium assumption used in
deriving the Michaelis-Menten equation \cite{segel} has been replaced 
with the notion of rapid
convergence of the probability distribution to a stationary one.
Note that we consider the behavior of a single enzyme molecule
rather than the collective behavior of the enzyme and enzyme-substrate complex.
In this way, a stochastic formulation of biological regulation gives an alternative interpretation of the Michaelis-Menten equation.
It should be noted that we have not assumed the enzyme concentration
is constant, which is required to derive
Michaelis-Menten equation in the traditional way.

It is not difficult to see that the standard deviation
$\sigma_{\upsilon}=\sqrt{\bar{\upsilon^{2}}-\bar{\upsilon}^{2}}$ is given by
\[
\sigma _{\upsilon}=k_{2}\frac{\sqrt{K_{1}\left[S \right]}}{K_{1}+\left[S \right]},
\]
which gives a rough estimate of the precision of the approximation (\ref{eq2-11}).
This is a bonus of the stochastic version of the Michaelis-Menten equation 
that we have derived.
\end{ex}

\begin{ex}\upshape
(Allosteric enzyme with an activator \cite{segel})\\
An allosteric enzyme $e$ with one binding site for
an activator $A$ in addition to the one for a substrate $S$ is described as
\begin{align}
\label{ex2-1}
&E\ + \ S\  \overset{k_{1}}{\underset{k_{-1}}{\rightleftarrows}} ES \overset{k_{p}}{\rightarrow}\ E\ +\ P \nonumber\\
&+ \hspace*{19mm} +   \nonumber\\
&A \hspace*{21mm} A   \\
{\scriptstyle k_{-2}}\!&\! \uparrow\downarrow \!{\scriptstyle k_{2}} \qquad\ \ {\scriptstyle k_{-3}}\! \uparrow\downarrow \!{\scriptstyle k_{3}}  
\nonumber\\
E&A\ +\ S\  \overset{k_{4}}{\underset{k_{-4}}{\rightleftarrows}}\ ESA \ \overset{k_{pA}}{\rightarrow}\ EA\ +\ P 
\nonumber
\end{align}
The enzyme has now the two binding sites, one for the substrate 
and the other for the activator $A$.
The enzyme has four states $S_{0}=(\phi\phi)$, $S_{1}=(S\phi)$,
$S_{2}=(SA)$, $S_{3}=(\phi A)$.
The state transition diagram is shown in Fig.~\ref{fig:1}(b).
We can associate the transition probabilities with 
the rate constants in (\ref{ex2-1}) as
\begin{align}
&q_{10}=k_{1}\left[S \right],\ q_{01}=k_{-1}, \ q_{21}=k_{3}\left[A \right],
\ q_{12}=k_{-3}, \nonumber \\
&q_{32}=k_{-4}, \ q_{23}=k_{4}\left[S \right],
\ q_{03}=k_{-2}, \ q_{30}=k_{2}\left[A \right],\label{eq2-13} 
\end{align}
where $\left[S \right]$ and $\left[A \right]$ denote the 
concentrations of the substrate and the activator, respectively.
The transition matrix is given by
\begin{equation}
Q=\begin{bmatrix}
D_{0}&q_{01}&0&q_{03}\\
q_{10}&D_{1}&q_{12}&0\\
0&q_{21}&D_{2}&q_{23}\\
q_{30}&0&q_{32}&D_{3}
\end{bmatrix}.
\end{equation}
where $D_{0}=-(q_{10}+q_{30})$, $D_{1}=-(q_{01}+q_{21})$,
$D_{2}=-(q_{12}+q_{32})$, $D_{3}=-(q_{03}+q_{23})$.
The rate of the target chemical reactions producing $P$ is given by
\[
\gamma =k_{p}p_{1}+k_{pA}p_{2}
\]

\noindent
The actual computation of the stationary probability distribution
for this example is done in Example 6, where you see the result is
very complex.
\end{ex}

\begin{ex}\upshape
(Allosteric enzyme with both activator and inhibitor)\\
The reaction scheme is written as
\begin{align}
\label{ex3-16}
&EI\ +\ S\ \overset{k_{7}}{\underset{k_{-7}}{\rightleftarrows}}\ ESI \ \overset{k_{pI}}{\rightarrow}\ EI\ +\ P \nonumber \\
{\scriptstyle k_{6}}\!&\! \uparrow\downarrow \!{\scriptstyle k_{-6}} \qquad\quad\ {\scriptstyle k_{5}}\! \uparrow\downarrow \!{\scriptstyle k_{-5}}  
\nonumber\\
&\ I \hspace*{24mm} I \nonumber \\
&+ \hspace*{22mm} + \nonumber \\
&E\ +\ S\ \ \overset{k_{1}}{\underset{k_{-1}}{\rightleftarrows}}\ \ ES \ \overset{k_{p}}{\rightarrow}\ E\ +\ P \\
&+ \hspace*{22mm} + \nonumber \\
&A \hspace*{24mm} A \nonumber \\
{\scriptstyle k_{-2}}\!&\! \uparrow\downarrow \!{\scriptstyle k_{2}} \qquad\quad\ {\scriptstyle k_{-3}}\! \uparrow\downarrow \!{\scriptstyle k_{3}}  
\nonumber\\
&\! EA\ +\ S \ \overset{k_{4}}{\underset{k_{-4}}{\rightleftarrows}}\ ESA \  \overset{k_{pA}}{\rightarrow}\ EA\ +\ P \nonumber
\end{align}
where I denotes the inhibitor.
The enzyme has six states, namely,
$S_{0}=(\phi\phi)$, $S_{1}=(\phi A)$, $S_{2}=(SA)$,
$S_{3}=(S \phi)$, $S_{4}=(SI)$, $S_{5}=(\phi I)$.
The state transition diagram is shown in Fig.~\ref{fig:1}(c).
The right part is identical to the diagram of Fig.~\ref{fig:1}(a).
The state transition matrix is given by
\begin{equation}
Q=\begin{bmatrix}
D_{0}&q_{01}&0&q_{03}&0&q_{05}\\
q_{10}&D_{1}&q_{12}&0&0&0\\
0&q_{21}&D_{2}&q_{23}&0&0\\
q_{30}&0&q_{32}&D_{3}&q_{34}&0\\
0&0&0&q_{43}&D_{4}&q_{45}\\
q_{50}&0&0&0&q_{54}&D_{5}
\end{bmatrix},
\end{equation}
where $D_{0}=-(q_{10}+q_{30}+q_{50})$, 
$D_{1}=-(q_{01}+q_{21})$, $D_{2}=-(q_{12}+q_{32})$,
$D_{3}=-(q_{03}+q_{23}+q_{43})$, $D_{4}=-(q_{34}+q_{54})$,
$D_{5}=-(q_{05}+q_{45})$.     
The rate of the target chemical reactions is given by
\[
\gamma=k_{pA}p_{2}+k_{p}p_{3}+k_{pI}p_{4}
\]
The computation of the stationary probability distribution is
done in Example 6.
\end{ex}

\begin{figure}[h]
  \begin{center}
    \includegraphics[width=0.3\textwidth]{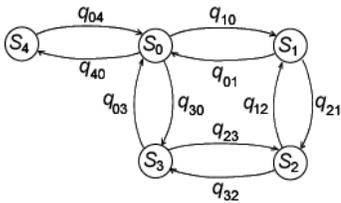}
  \end{center}
  \caption{State transition diagram of {\it lac} operon.}
  \label{fig:2}
\end{figure}

\begin{ex}\upshape
({\it Lac} operon)\\
The {\it lac} operon, which has been a subject of
genetic molecular biology for more than fifty years 
(\cite{bliss}\cite{santillan}\cite{wong}\cite{Zaslaver}), can be 
regarded as a biological regulator. 
A number of different models have been proposed for it including
a very complicated model with seven \textit{cis}-regulatory elements \cite{santillan}.
The standard model has three binding sites \cite{Zaslaver}.
One is the promoter with RNAP as its unique binding factor.

There are two binding factors associated with the {\it lac} operon,
cAMP-CRP and LacI, which have their own 
binding sites. They are independent of each other.
cAMP-CRP is an activator of the operon, while LacI
is a repressor.
Binding of LacI at its site prevents the other factors from 
binding to their sites.

Denote the RNAP, cAMP-CRP and LacI by $U_{1}$, $U_{2}$ and $U_{3}$, respectively. 
Then, the {\it lac} operon has the following five states;
\begin{align*}
&S_{0}=(\phi\phi\phi),\ S_{1}=(U_{1}\phi\phi),\ S_{2}=(U_{1}U_{2}\phi),\\ 
&S_{3}=(\phi U_{2}\phi),\ S_{4}=(\phi\phi U_{3}). 
\end{align*} 
The transition diagram is shown in Fig.~\ref{fig:2}.  
The transcription rate is given by
\[ \gamma=\gamma_{1}p_{1}+\gamma_{2}p_{2}, \]
where $\gamma_{1}$ and $\gamma_{2}$ are the transcription rates corresponding to $S_{1}$ and $S_{2}$.
\end{ex}

\begin{figure}[h]
  \begin{center}
    \includegraphics[width=0.5\textwidth]{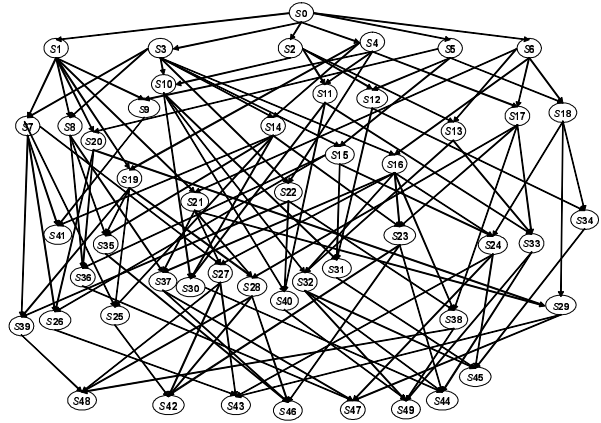}
  \end{center}
  \caption{Transition diagram of {\it lac} operon with 50 states.}
  \label{fig:3}
\end{figure}

Recently, Santillan et al. proposed a lac operon model with 
five binding sites \cite{santillan}.
The number of binding patterns of its \textit{cis}-regulatory units is
at least 50.
Figure~\ref{fig:3} shows the transition diagram which is very complicated.
We need an efficient computational theory to handle complex operons
like this.

\section{Conservation of probability flows in stationary distribution
and loop-free transition diagram}
\label{sec3}

\begin{figure}[h]
  \begin{center}
    \includegraphics[width=0.3\textwidth]{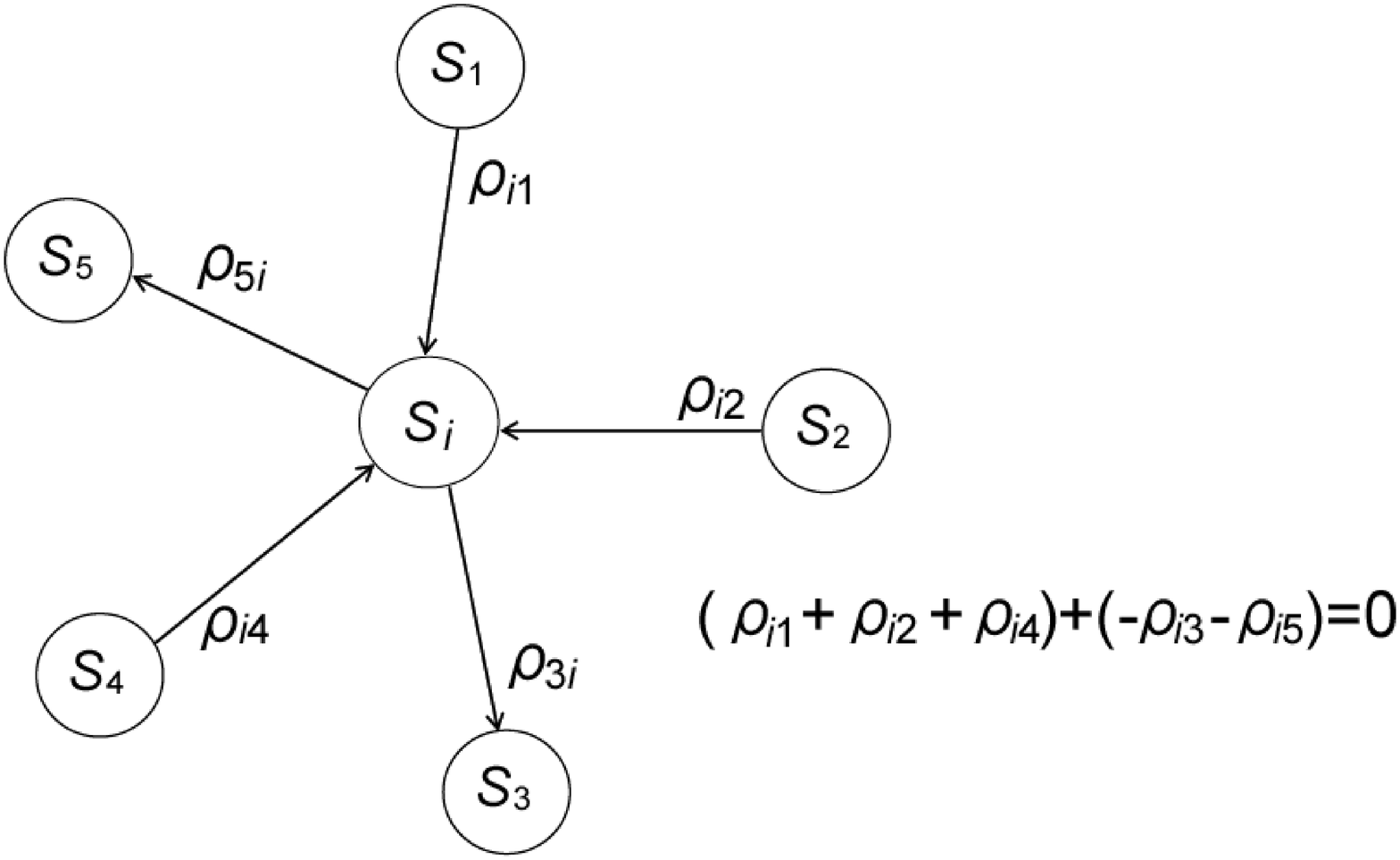}
  \end{center}
  \caption{Conservation of probability flows.}
  \label{fig:4}
\end{figure}

In this section, we introduce the notion of probability flow and discuss its meaning
in solving the SSE (\ref{eq2-4a}).
The SSE (\ref{eq2-4a}) can be written as
\begin{equation}
\sum_{j\in {\bf A}_{i}}(q_{ij}p_{j}-q_{ji}p_{i})=0,\quad i=0,1,\cdots,N.
\label{eq8}
\end{equation}
The term $q_{ij}p_{j}-q_{ji}p_{i}$ represents the net probability
of the state transition from $S_{j}\in {\bf A}_{i}$ to $S_{i}$.
Hence, it is reasonable to call it the \textit{probability flow from
$S_{j}$ to $S_{i}$}, which is denoted by
\begin{equation}
\rho_{ij}=q_{ij}p_{j}-q_{ji}p_{i}.
\label{eq9}
\end{equation}
If $\rho_{ij}>0$, the probability flows from $S_{j}$ to $S_{i}$
and if $\rho_{ij}<0$, it flows oppositely.
Clearly, it is \textit{skew-symmetric}, i.e.,
\begin{equation}
\rho_{ij}+\rho_{ji}=0.
\label{eq10}
\end{equation}
The SSE (\ref{eq8}) is represented in terms of probability flow as
\begin{equation}
\sum_{j\in {\bf A}_{i}}\rho_{ij}=0, \quad \forall i,
\label{eq11}
\end{equation}
which implies that the probability flows are conserved 
at each state node.
In other words, the net probability flow incoming to $S_{i}$
(the sum of positive $\rho _{ij}$) is equal to the outgoing
flow from $S_{i}$ (the sum of negative $\rho_{ij}$), as
is illustrated in Figure~\ref{fig:4}.
Here, it is important to notice that the probability flow is directed.

\begin{figure}[h]
  \begin{center}
    \includegraphics[width=0.35\textwidth]{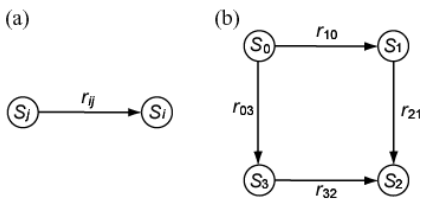}
  \end{center}
  \caption{TR Diagrams.
(a) Description of TR, (b) TR diagram of Fig.~\ref{fig:1}(b).}
  \label{fig:5}
\end{figure}

In chemical kinetics, one usually assumes that the equilibrium state
satisfies the relations
\begin{equation}
q_{ij}p_{j}=q_{ji}p_{i},\quad \forall j\in {\bf A}_{i},\quad \forall i,
\label{eq12}
\end{equation}
if we interpret $q_{ij}$ as the kinetic rate coefficient of the reaction
$S_{j}\to S_{i}$.
The relations (\ref{eq12}) are called \textit{detailed balance} in 
chemical kinetics \cite{heinrich}.
Due to (\ref{eq9}), the detailed balance holds if and only if
$\rho_{ij}=0$, $\forall i$, $j\in {\bf A}_{i}$; i.e., all the probability
flows vanish.
The relation (\ref{eq12}) is written as 
\begin{equation}
p_{i}=r_{ij}p_{j}
\label{eq13a}
\end{equation}
where $r_{ij}$ is called the \textit{transition ratio} from
$S_{j}$ to $S_{i}$ and is defined as 
\begin{equation}
r_{ij}=\frac{q_{ij}}{q_{ji}}.
\label{eq14a}
\end{equation}
The transition ratio (TR) is directed as shown in Fig.~\ref{fig:5}(a)
and corresponds to the equilibrium coefficient of the chemical reaction
$S_{j}\leftrightarrows S_{i}$.
It is sometimes convenient to describe the transition
diagram in terms of the transition ratios (TR), instead of 
transition probabilities, as is shown in Fig.~\ref{fig:5}(b).
We call such a diagram a \textit{TR diagram} to distinguish it from the usual transition 
diagram. Note that
\begin{equation}
r_{ij}=r^{-1}_{ji}.
\label{e24q}
\end{equation}

\begin{figure}[h]
  \begin{center}
    \includegraphics[width=0.4\textwidth]{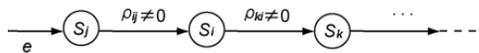}
  \end{center}
  \caption{Sequence of edges with non-zero probability flow.}
  \label{fig:6}
\end{figure}

\noindent
The direction of an edge can be freely assigned in TR diagrams,
but it must be consistent with the direction of the TR,
in the sense that an edge with $r_{ij}$ as its TR must be directed
from $S_{j}$ to $S_{i}$.
Now, we shall prove that the probability flows vanish at all
edges unless the transition diagram has a loop.
Here, a loop is defined in the context of the TR diagram.
In other words, if the transition diagram does not have a loop
(loop-free), all the probability flows vanish.
To see this remarkable fact, assume that there exists
an edge $e$ with a non-zero probability flow and a state $S_{j}$ is
connected to this edge.
Then, due to the conservation of probability flow at $S_{j}$, there exists
another edge connecting $S_{j}$ to another state $S_{i}$ with
a non-zero probability flow.
The same reasoning for $S_{i}$ leads one to conclude that $S_{i}$ must
be connected to a new state $S_{k}$ by an edge with a
non-zero probability flow (Fig.~\ref{fig:6}).
Repeating the procedure creates a sequence of states
$S_{j}\to S_{i}\to S_{k}\to \cdots $, which are connected by
edges with non-zero probability flows.
Since the number of states is finite, the sequence must visit
a state which has already appeared in the sequence.
Hence, the graph must contain a loop.

For a while, we concentrate on the loop-free case
where all the probability flows vanish.
Take two arbitrary states $S_{i}$ and $S_{j}$.
If there are two different paths connecting $S_{i}$ and $S_{j}$,
it means that there is a loop composed of a path from $S_{j}$ to
$S_{i}$ and one from $S_{i}$ to $S_{j}$.
Therefore, if the transition diagram is loop-free, the state
$S_{i}$ is connected to $S_{j}$ through a unique path.
Let this path be composed of $l$ state: 
$S_{j}=S_{i_{1}}\to S_{i_{2}}\to \cdots \to S_{i_{l}}=S_{i}$.
We can define the transition probability from $S_{j}$ to $S_{i}$ as
\begin{equation}
\bar{q}_{ij}=q_{ii_{l-1}}q_{i_{l-1}i_{l-2}}\cdots q_{i_{2}j},
\label{eq13}
\end{equation} 
where the overbar is used to distinguish the transition probability
computed along a path from those between adjacent nodes.
In this way, we can compute the transition probability for any state
pair $S_{i}$ and $S_{j}$ which are not necessarily adjacent to each other.

We can extend the detailed balance (\ref{eq12}) to any pair of states
by using the extended transition probabilities (\ref{eq13}).

To see this, assume that three states $S_{0}$, $S_{1}$, and $S_{2}$ are
connected by a path, i.e., $S_{1}\in {\bf A}_{0}$ and $S_{2} \in {\bf A}_{1}$.
Then, the detailed balance (\ref{eq12}) implies
$q_{10}p_{0}=q_{01}p_{1}$ and $q_{21}p_{1}=q_{12}p_{2}$.
Multiplying each side yields 
$q_{10}p_{0}q_{21}p_{1}=q_{01}p_{1}q_{12}p_{2}$.
Cancelling $p_{1}$ from both sides gives $q_{21}q_{10}p_{0}=q_{01}q_{12}p_{2}$,
or equivalently, $\bar{q}_{20}p_{0}=\bar{q}_{02}p_{2}$.
This relation suggests that the detailed balance holds 
in terms of the generalized transition probability (\ref{eq13}) even
for state pairs which are not adjacent to each other. 
It is not difficult to see that this is indeed the case
by repeatedly applying it through the unique
path connecting the two states, i.e.,
\begin{equation}
\bar{q}_{ij}p_{j}=\bar{q}_{ji}p_{i},\quad \forall i,j.
\label{eq14}
\end{equation}
The above relation includes the detailed balance (\ref{eq12})
as a special case.
Therefore, we call the relation (\ref{eq14}) the
\textit{generalized detailed balance}.
Taking $j=0$ in (\ref{eq14}), we can represent $p_{j}$ as
\begin{equation}
p_{i}=\bar{r}_{i0}p_{0},
\label{eq15}
\end{equation}
where $\bar{r}_{ij}$ is defined as a generalization of (\ref{eq14a}) with
$q_{ij}$ being replaced by $\bar{q}_{ij}$ given by (\ref{eq13}), i.e.,
\begin{equation}
\bar{r}_{ij}=\frac{\bar{q}_{ij}}{\bar{q}_{ji}}=r_{ii_{l-1}}r_{i_{l-1}i_{l-2}}\cdots r_{i_{2}j}
\label{eq16}
\end{equation}
Here, the overbar is again used to distinguish it from $r_{ij}$ given
in (\ref{eq14a}) for state pairs adjacent to each other.
$\bar{r}_{ij}$ is also called the transition ratio (TR) from $S_{j}$ to $S_{i}$. 
Note that the relation (\ref{e24q}) is extended to 
\[
\bar{r}_{ij}=\bar{r}^{-1}_{ji}.
\]
From the normalization constraint (\ref{eq2-4}) and (\ref{eq15}), 
the statitonary probability distribution is
simply given by
\begin{equation}
p_{i}=\frac{\bar{r}_{i0}}{\sum^{N}_{j=0}\bar{r}_{j0}},\quad
i=0,1,\cdots,N
\label{eq17}
\end{equation}
where $\bar{r}_{00}=1$.

We sum up the discussion in this section as follows:\\
\textit{If the transition diagram is loop-free, 
the following facts hold: \\
(1) the probability flows vanish at every edge. \\
(2) The stationary distribution is given by (\ref{eq17}).\\
(3) The generalized detailed balance (\ref{eq14}) holds for
the stationary distribution.}

\begin{ex}\upshape
(Loop-free Transition Diagram)\\
Consider a transition diagram of Fig.~\ref{fig:7}, which is loop-free.
According to (\ref{eq15}), we have $p_{1}=q_{10}p_{0}$, 
$p_{2}=r_{21}r_{10}p_{0}$, $p_{3}=r_{31}r_{10}p_{0}$,
$p_{4}=r_{43}r_{31}p_{0}$, $p_{5}=r_{50}p_{0}$ with
\[
p_{0}=\frac{1}{1+r_{10}(1+r_{21}+r_{31}+r_{43}r_{31})+r_{50}}
\]

\begin{figure}[h]
  \begin{center}
    \includegraphics[width=0.3\textwidth]{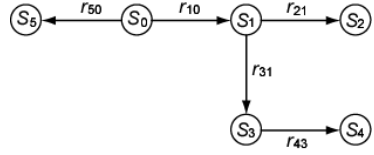}
  \end{center}
  \caption{A loop-free transition diagram.}
  \label{fig:7}
\end{figure}

The above reasoning can be applied to a loop-free subgraph of
the general (not loop-free) transition diagram.
If a loop-free subgraph is attached to a state $S_{i}$ which 
belongs to a loop, we can write down the probability of the state
of that subgraph according to (\ref{eq17}) with $p_{0}$ being replaced by $p_{i}$.
As an example, consider the transition diagram of Fig.~\ref{fig:8}.
The state $S_{j}$ is a part of a loop which will be discussed
in the next section.
If the state $S_{i}$ is not in any loop, then it is a part of a path
connecting $S_{i}$ and the terminal state $S_{k}$. Thus,
\begin{equation}
p_{i}=\bar{r}_{ij}p_{j},
\label{eq31new}
\end{equation}
because the probability flow at any edge on the path connecting
$S_{j}$ and $S_{k}$ vanishes.
This can be shown by directly applying the argument of this section
to the subgraph of Fig.~\ref{fig:8}.
\begin{figure}[h]
  \begin{center}
    \includegraphics[width=0.4\textwidth]{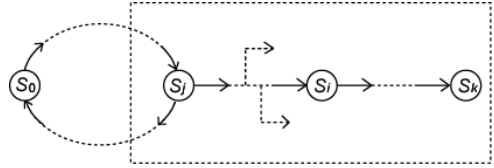}
  \end{center}
  \caption{A loop-free subgraph.}
  \label{fig:8}
\end{figure}

\end{ex}

\section{Wegscheider condition}
\label{sec4}

In the preceding section, we showed that the probability
flows vanish if the transition diagram is loop-free.
In that case, the stationary distribution  is simply calculated from (\ref{eq17}).
There are cases where the probability flows vanish 
even if the transition diagram contains loops.

\begin{figure}[t]
  \begin{center}
    \includegraphics[width=0.4\textwidth]{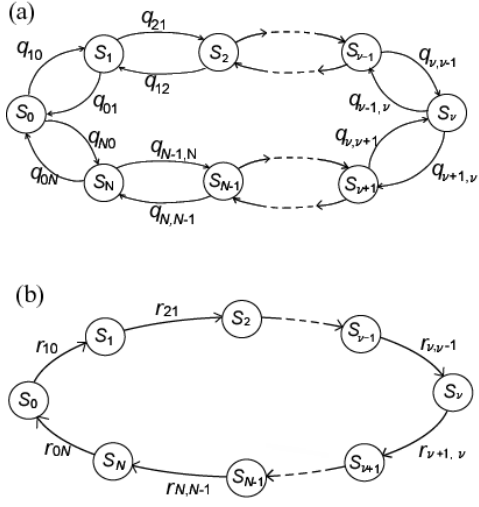}
  \end{center}
  \caption{Loop transition Diagram.
(a) Transition probability description, (b) transition ratio description.}
  \label{fig:9}
\end{figure}

In order to discuss this issue, we consider the case 
where the whole transition diagram is a loop, as shown in
Fig.~\ref{fig:9}(a).
The corresponding TR diagram is shown in Fig.~\ref{fig:9}(b).
The state transition matrix is given by
\begin{equation}
\label{eq5-1eq}
Q=\begin{bmatrix}D_{0}&q_{01}&0&\cdots&0&q_{0N}\\
q_{10}&D_{1}&q_{12}&\cdots&0&0\\
0&q_{21}&D_{2}&\cdots&0&0\\
 &\cdots&\cdots&&&\\
 q_{N0}&0&0&\cdots&q_{N,N-1}&D_{N}
 \end{bmatrix}
\end{equation} 
with $D_{i}=-(q_{i+1,i}+q_{i-1,i})$, $i=1,2,\cdots,N-1$,
$D_{0}=-(q_{10}+q_{N0})$, $D_{N}=-(q_{0N}+q_{N-1,N})$.

Each state has only two edges, one of which
receives the probability flow, the other dispatches it. 
Due to the conservation of probability flows (\ref{eq11}),
the incoming probability flow and the outgoing one are equal.
Hence, each edge of the loop has the same amount of probability flow,
which is denoted by $\rho$.
Taking its direction to be clockwise, that is, taking the
probability flow from $S_{i-1}$ to $S_{i}$ to be positive, 
the probability flow is given by
\begin{equation}
\rho=q_{i,i-1}p_{i-1}-q_{i-1,i}p_{i},\quad i=1,2,\cdots,N.
\label{5-eq29}
\end{equation}
Thus, we have the following recursion:
\begin{align}
\label{eq18}
p_{i}&=r_{i,i-1}p_{i-1}-\frac{\rho}{q_{i-1,i}},\quad
i=1,2,\cdots,N, \\
p_{0}&=r_{0N}p_{N}-\frac{\rho}{q_{N,0}}. \nonumber
\end{align}

In order to obtain a general representation of stationary
probabilities, we introduce the numbers $\xi_{0},\xi_{1},\cdots,\xi_{N}$ 
defined sequentially by
\begin{align}
\xi_{0}&=0 \nonumber\\
\xi_{i}&=r_{i,i-1}\xi_{i-1}-\frac{1}{q_{i-1,i}},\quad
i=1,2,\cdots,N 
\label{eq23}
\end{align}
Then, due to (\ref{eq18}) and (\ref{eq23}), 
$p_{i}=r_{i,i-1}p_{i-1}+\rho(\xi_{i}-r_{i,i-1}\xi_{i-1})$, 
Therefore, we have
\[ p_{i}-\rho\xi_{i}=r_{i,i-1}(p_{i-1}-\rho\xi_{i-1}), \]
which implies $p_{i}-\rho\xi_{i}=\bar{r}_{i0}p_{0}$.
Here, $\bar{r}_{i0}=r_{i,i-1}r_{i-1,i-2}\cdots r_{10}$.
Thus, $p_{i}$ is represented as
\begin{equation}
p_{i}=\bar{r}_{i0}p_{0}+\rho \xi_{i}, \quad i=1,\cdots,N.
\label{eq24}
\end{equation}

If $\rho=0$, the above relations are identical to (\ref{eq15}).
Thus, equation (\ref{eq24}) represents the stationary 
distribution as a sum of the probability distribution for the case
with zero probability flow and a correction term due to 
non-zero probability flow.
Since $\rho=q_{0N}p_{N}-q_{N0}p_{0}$, the relation (\ref{eq24})
for $i=N$ implies $\rho=q_{0N}(\bar{r}_{N0}p_{0}+\rho\xi_{N})-q_{N0}p_{0}=q_{N0}(r_{0N}\bar{r}_{N0}-1)p_{0}+\rho q_{0N}\xi_{N}$.
Therefore, we get
\begin{equation}
\rho(1-q_{0N}\xi_{N})=-(1-\eta)q_{N0}p_{0}
\label{eq35eq}
\end{equation}
where $\eta$ is given by
\begin{equation}
\eta=r_{0N}\bar{r}_{N0}=r_{0N}r_{N,N-1}\cdots r_{10}
\label{eq36eq}
\end{equation}
Since $\xi_{1}=-1/q_{01}<0$ and $r_{ij}>0$, $\xi_{N}<0$.
Hence, $1-q_{0N}\xi_{N}>0$. Therefore, we conclude that 
the probability flow vanishes, i.e., $\rho=0$, if and only if $\eta=1$. 

The number $\eta$ is the product of all the
TRs along the loop. If we use the analogy between the TR and
the equilibrium coefficient in chemical kinetics, the condition
$\eta=1$, i.e.,
\begin{equation}
r_{10}r_{21}\cdots r_{0N}=1,
\label{eq37eq}
\end{equation}
corresponds to the condition derived almost a century ago
by Wegscheider. The condition (\ref{eq37eq}) is referred to
as the \textit{Wegscheider condition} \cite{karmarkar}.
We call the product $\eta$ of (\ref{eq36eq})
the \textit{Wegscheider product} for a loop as in Fig.~\ref{fig:9}.
The Wegscheider product is in some sense directed.
The representation (\ref{eq36eq}) defines the Wegscheider product
clockwise in the context of Fig.~\ref{fig:9}.
It can also be defined in the opposite direction
$\eta'=r_{N0}r_{N-1,N}\cdots r_{01}$.
Due to (\ref{e24q}), $\eta'=\eta^{-1}$.
Since $\eta=1$ implies $\eta'=1$, the Wegscheider condition can be
represented by computing the product in either clockwise 
or counter clockwise direction.

The Wegscheider condition is a property of a loop that
requires the transition ratio along the total loop to be unity.
This condition is equivalently written in terms of transition
probabilities as
\begin{equation}
q_{10}q_{21}\cdots q_{0N}=q_{N0}q_{N-1,N}\cdots q_{01}.
\label{eq21}
\end{equation}
The left-hand side represents the transition probability from
a state to itself along a clockwise contour, while 
the right-hand side represents that along a counter clockwise contour.
The identity (\ref{eq21}) gives a stochastic interpretation of the Wegscheider condition.

Taking the normalization condition (\ref{eq2-4}) into account, we have
\begin{equation} 
p_{i}=\frac{\bar{r}_{i0}+\rho\xi_{i}}{\sum^{N}_{j=0}\bigl(\bar{r}_{j0}+\rho\xi_{j} \bigr)},
\quad i=0,1,\cdots,N,
\label{eq27}
\end{equation}
where $r_{00}=1$, $\xi_{0}=0$.
If the Wegscheider condition holds, then $\rho=0$ and the stationary
distribution becomes identical to (\ref{eq17}).

The most common method of quantifying the regulatory activities of operons
is based on thermal equilibrium theory \cite{berg}\cite{bintu}, which assigns a Gibbs free energy to each binding pattern of the binding sites.
This corresponds to assigning a stationary probability distribution
a priori, rather than constructing a Markov process by assigning
the transition probabilities between states.
The transition ratio may then be defined as the ratio between the state probabilities.
In that case, the Wegscheider condition obviously holds.
To see this, consider a loop $S_{0}\to S_{1}\to S_{2}\to S_{0}$.
Then, $r_{10}=p_{1}/p_{0}$, $r_{21}=p_{2}/p_{1}$, $r_{02}=p_{0}/p_{2}$.
Thus, $r_{10}r_{21}r_{02}=1$.
This proves that the Wegscheider condition is consistent with thermal 
equilibrium theory.
The relation (\ref{eq27}), however, suggests more general types of
equilibrium.

\section{Edge removal and modifier}
\label{sec5}

\begin{figure}[h]
  \begin{center}
    \includegraphics[width=0.5\textwidth]{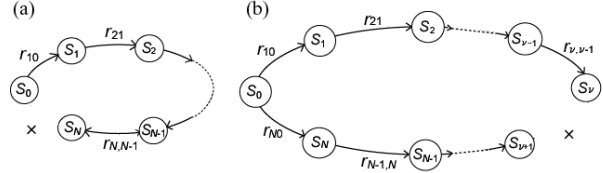}
  \end{center}
  \caption{Reduced transition diagrams.
(a) Edge $S_{N}\to S_{0}$ is eliminated, (b) edge $S_{\nu}\to S_{\nu+1}$ is eliminated.}
  \label{fig:10}
\end{figure}

In this section, we consider the meaning of the numbers $\xi_{i}$
in (\ref{eq23}) and try to interpret (\ref{eq24}) in the 
context of edge removal.
From (\ref{eq23}), $q_{i-1,i}\xi_{i}-q_{i,i-1}\xi_{i-1}=-1$ and
$q_{i,i+1}\xi_{i+1}-q_{i+1,i}\xi_{i}=-1$.
These relations yield $-q_{i,i-1}\xi_{i-1}-q_{i,i+1}\xi_{i+1}+(q_{i+1,i}+q_{i-1,i})\xi_{i}=0$.
Since $\xi_{0}=0$ and $\xi_{1}=-1/q_{01}$, we have, from (\ref{eq5-1eq}),
\begin{equation}
Q\xi=\begin{bmatrix}-1\\0\\\vdots\\0\\1\end{bmatrix}(1-q_{0N}\xi_{N}),
\label{eq38eq}
\end{equation}
where $\xi=\begin{bmatrix}\xi_{0}&\xi_{1}&\cdots&\xi_{N}\end{bmatrix}^{T}$.
To further investigate the meaning of the vector $\xi$,
we consider a transition diagram which is obtained from the
original diagram of Fig.~\ref{fig:9} by eliminating the edge
connecting $S_{0}$ and $S_{N}$ (Fig.~\ref{fig:10}(a)).
The modified transition diagram has no loop, and
hence, the stationary distribution is given by $\bar{p}_{i}=\bar{r}_{i0}p_{0}$,
following the discussion in the preceding section.
The transition matrix $Q'$ corresponding to the modified diagram
of Fig.~\ref{fig:10}(a) is obtained by taking $q_{0N}=q_{N0}=0$ in 
(\ref{eq5-1eq}).
From (\ref{eq38eq}), we have
\begin{equation}
Q'\xi=\begin{bmatrix}-1\\0\\ \vdots\\0\\1\end{bmatrix}.
\label{eq39eq}
\end{equation}
We call the vector $\xi$ satisfying (\ref{eq39eq}) with $\xi_{0}=0$
\textit{the modifier} of the loop corresponding to the edge $S_{N}\to S_{0}$.
It is computed through the recursion formula (\ref{eq23}).
We shall generalize it in the sequel.

If we let $p'$ be the vector whose $(j+1)$th element $p_{i}$
is given by (\ref{eq15}), which represents
the stationary probability distribution corresponding to
the modified loop-free diagram of Fig.~\ref{fig:10}(a),
equation (\ref{eq24}) can be written as
\begin{equation}
p=p'+\rho \xi.
\label{eq40eq}
\end{equation}
This is the general form of the stationary probability
distribution. It holds for a loop transition diagram as well as for general 
transition diagram with multiple loops, as is discussed in the next section.
Note that the stationary distribution is represented by a 
stationary distribution $p'$ corresponding to the modified loop-free
diagram with a correcting term $\rho\xi$ due to non-zero probability flow.

Equation (\ref{eq40eq}) is obtained by cutting the edge $S_{N}\to S_{0}$
as in Fig.~\ref{fig:10}(a).
Here, we can show that equation (\ref{eq40eq}) is also
derived by eliminating any edge $S_{\nu}\to S_{\nu+1}$ of
the loop (Fig.~\ref{fig:10}(b)).
Since $\bar{r}_{i0}$ is defined as the product of each TR
along the path connecting $S_{i}$ to $S_{0}$, we have
\begin{equation}
\bar{r}_{i0}=\left\{%
\begin{array}{lll}
r_{10}r_{21}\cdots r_{i,i-1} & \mbox{if} & i\le \nu. \\
r_{N0}r_{N-1,N}\cdots r_{i,i-1} & \mbox{if} & i\ge \nu+1.
\end{array}\right.
\label{eq41eq}
\end{equation}
Now, the modifier is defined as
\begin{align}
&\mbox{[\textit{Forward Recursion}]}\nonumber\\
&\xi_{0}=0 \nonumber \\
&\xi_{i}=r_{i,i-1}\xi_{i-1}-\frac{1}{q_{i-1,i}}, \quad i=1,2,\cdots,\nu
\label{eq42eq}\\
&\mbox{[\textit{Backward Recursion}]}\nonumber\\
&\xi_{N+1}=0. \quad q_{N+1,N}=q_{0N} \nonumber \\
&\xi_{i}=r_{i,i+1}\xi_{i+1}+\frac{1}{q_{i+1,i}}, \quad i=N,N-1,\cdots,\nu+1. 
\label{eq44plus}
\end{align}
The recursion formula (\ref{eq42eq}) computes the modifier elements of
the path connecting $S_{0}$ to $S_{\nu}$, while the recursion formula
(\ref{eq44plus}) computes those of the path connecting $S_{0}$ to
$S_{\nu+1}$.
The paths are oppositely directed, but both recursions are
essentially the same except the signs of the added term.
We call (\ref{eq42eq}) \textit{forward recursion}, while (\ref{eq44plus})
\textit{backward recursion}.
It is straightforward to see that 
$\xi=\begin{bmatrix}\xi_{0}&\xi_{1}&\cdots&\xi_{N}\end{bmatrix}^{T}$
satisfies
\[
Q\xi=\begin{bmatrix}0&0\\\vdots&\vdots\\0&0\\-q_{\nu+1,\nu}&q_{\nu,\nu+1}\\
q_{\nu+1,\nu}&-q_{\nu,\nu+1}\\0&0\\\vdots&\vdots\\0&0\end{bmatrix}
\begin{bmatrix}\xi_{\nu}\\\xi_{\nu+1}\end{bmatrix}+\begin{bmatrix}0\\\vdots\\0\\1\\-1\\0\\\vdots\\0\end{bmatrix}.
\]
Since $q_{\nu+1,\nu}=q_{\nu,\nu+1}=0$ in the modified transition
diagram, we have
\begin{equation}
\begin{array}{rcll}
\ldelim[{6}{0.1mm}[]&0& \rdelim]{6}{0.1mm}[]\\
                    &\vdots&                \\
       Q'\xi=       &1&  &\cdots   \nu+1 \\
                    &-1& &\cdots   \nu+2  \\
                    &\vdots&  \\
                    &0& \\
\end{array}
\label{eq44eq}
\end{equation}
It is not difficult to see that the stationary distribution is given by
(\ref{eq24}) with $\bar{r}_{i0}$ and $\xi_{i}$ being represented by
(\ref{eq41eq}) and (\ref{eq42eq})(\ref{eq44plus}), respectively,
by applying
the arguments of the  preceding section to the forward and
backward recursions separately.


The modifier is extended to any edge $S_{\nu}\to S_{\nu+1}$ and can be
computed through (\ref{eq42eq})(\ref{eq44plus}).
The stationary probability distribution is given in this case by (\ref{eq40eq}), 
where $p'$ corresponds to the stationary probability
distribution of the reduced loop-free diagram.
Since $\rho=0$ holds under Wegscheider condition, we obtain the following result: 

\textit{The stationary distribution of a loop is unchanged
if an edge is removed, provided that the Wegscheider condition
(\ref{eq37eq}) or (\ref{eq21}) holds}. \\
In other words, the computation of the stationary distribution
for a loop transition diagram is reduced to the loop-free case
where a very simple form (\ref{eq17}) is already available by eliminating
an edge, provided that the Wegsheider condition holds.
We shall extend this remarkable property to the general transition
diagram with multiple loops in the next section.


\begin{ex}\upshape
(Stationary Distribution of Example 2)\\
This is the case of $N=3$ in the above algorithm.
Remove the edge $e: S_{2}\to S_{3} (\nu=2)$.
The numbers $\xi_{i},i=0,1,2,3,$ are given respectively by
forward recursion
\begin{align}
\xi_{0}&=0, \ \xi_{1}=-\frac{1}{q_{01}},\nonumber \\
\xi_{2}&=-\frac{r_{21}}{q_{01}}-\frac{1}{q_{12}}=-\frac{q_{21}+q_{01}}{q_{01}q_{12}} \label{eq:ex6-48}
\end{align}
and by backward recursion
\[
\xi_{3}=\frac{1}{q_{03}}
\]
The stationary probability distribution is calculated to be
\begin{align}
\label{eq5-1}
p_{1}&=r_{10}p_{0}-\frac{1}{q_{01}}\rho  \nonumber\\
p_{2}&=\bar{r}_{20}p_{0}-\rho\xi_{2}=r_{21}r_{10}p_{0}-\frac{q_{01}+q_{21}}{q_{01}q_{12}}\rho \\
p_{3}&=r_{30}p_{0}+\frac{1}{q_{03}}\rho. \nonumber
\end{align}
The Wegscheider product $\eta$ is given by
\[ \eta=r_{03}r_{32}r_{21}r_{10}=\frac{q_{03}q_{32}q_{21}q_{10}}{q_{30}q_{23}q_{12}q_{01}}. \]
The probability flow $\rho=q_{32}p_{2}-q_{23}p_{3}$ is given by 
\[ \rho=\frac{q_{03}q_{32}q_{21}q_{10}-q_{30}q_{01}q_{12}q_{23}}{q_{01}q_{12}(q_{03}+q_{32})+q_{03}q_{23}(q_{01}+q_{21})}p_{0} \]
\end{ex}

\section{Eliminating loops from the transition diagram and derivation of 
the flow equation}
\label{sec6}

\begin{figure}[h]
  \begin{center}
    \includegraphics[width=0.3\textwidth]{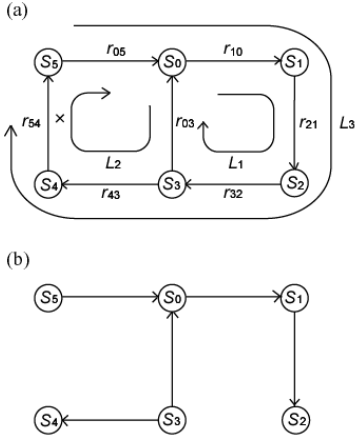}
  \end{center}
  \caption{An example with three loops.
(a) original diagram, (b) modified diagram.}
  \label{fig:11}
\end{figure}

Now, we consider the general case where multiple loops exist,
and derive a method of computing probability flows.
If some edges are removed, the transition diagram becomes loop-free.
Although the selection of such loops is not unique, 
the minimum number of such edges required to make the transition
diagram loop-free is unique.
As an example, consider a TR diagram of Fig.~\ref{fig:11}(a).
The elimination of the edges $e_{1}: S_{2}\to S_{3}$ and 
$e_{2}: S_{4}\to S_{5}$ makes the diagram loop-free, as is 
illustrated in Fig.~\ref{fig:11}(b).
We can eliminate, say, the edges $S_{0}\to S_{1}$ and $S_{3}\to S_{4}$
to make the diagram loop-free.
The elimination of two edges is necessary and
sufficient to make the diagram loop-free in this case.

Now, we recover the eliminated edges in the reduced loop-free 
diagram one by one. At each recovery step,
at least one new loop is recovered and we select one such loop.
These loops constitute \textit{basis loops}.
In case of Fig.~\ref{fig:11}, the recovery of edge
$e_{1}: S_{2}\to S_{3}$ recovers the loop $L_{1}: S_{0}\to S_{1}\to S_{2}\to S_{3}\to S_{0}$.
Then, recovery of $e_{2}: S_{4}\to S_{5}$ recovers 
$L_{2}: S_{0}\to S_{3}\to S_{4}\to S_{5}\to S_{0}$ and
$L_{3}: S_{0}\to S_{1}\to S_{2}\to S_{3}\to S_{4}\to S_{5}\to S_{0}$.
Thus, we can choose $L_{1}$ and $L_{2}$ as basis loops.
$L_{1}$ and $L_{3}$ can also be basis loops.
In this respect, each edge selected to make the transition
diagram loop-free corresponds to one basis loop.

Now, let us consider an algebraic representation for edge removal.
Consider an edge $e: S_{j}\to S_{i}$.
The probability flow of the edge $e$ from $S_{j}$ to $S_{i}$
is given by (\ref{eq9}), which is alternatively represented as
\begin{equation}
\rho_{ij}=\sigma (i,j)^{T}p,
\label{eq28}
\end{equation}
where $\sigma (i,j)$ is an $(N+1)$-vector given by
\begin{equation}
\left[\sigma(i,j)\right]_{k}=\left\{%
\begin{array}{lll}
q_{ij}, & k=j+1 \\
-q_{ji}, & k=i+1 \\
0, & \mbox{otherwise.}
\end{array}\right.
\label{eq29}
\end{equation} 
Note that $p_{i}$ appears at the $(i+1)$-th place 
instead of at the $i$-th place in the vector $p$ because of 
the existence of $p_{0}$ as its first element.
Removal of $e$ from the transition diagram
amounts to removal of $\rho_{ij}$ and $\rho_{ji}=-\rho_{ij}$
from the $(i+1)$-th and the $(j+1)$-th components of $Qp$,
respectively.
Therefore, removal of $e$ corresponds to removal of 
the matrix
\begin{equation}
\Delta Q:=-\delta(i,j)\sigma(i,j)^{T},
\label{eq30}
\end{equation}
where $\delta(i,j)$ is an $(N+1)$-dimensional vector
given by
\begin{equation}
\left[\delta(i,j)\right]_{k}=\left\{%
\begin{array}{lll}
-1, & k=i+1 \\
1, & k=j+1 \\
0 &
\end{array}\right.
\label{eq31}
\end{equation}
The reduced transition matrix $Q$ is given by
\begin{equation} 
Q'=Q-\Delta Q.
\label{eq32}
\end{equation}
Hence, from (\ref{eq30}) and (\ref{eq28}), we have
\[
Q'p=\bigl(Q+\delta (i,j)\sigma(i,j)^{T}\bigr)p=\rho_{ij}\delta(i,j).
\]
Assume that the transition diagram contains $l$ basis loops
$L_{1},L_{2},\cdots,L_{l}$. 
We can assume that all these loops contain $p_{0}$
without loss of generality.
Now, choose a set of edges $e_{k}\in L_{k},k=1,2,\cdots,l$,
such that the elimination of $e_{1},e_{2}\cdots,e_{l}$ makes
the transition diagram loop-free.
We assume that the edge $e_{k}$ connects the states
$S_{\mu_{k}}$ and $S_{\nu_{k}}$ in the direction from
$S_{\mu_{k}}$ to $S_{\nu_{k}}$, i.e., $e_{k}: S_{\mu_{k}}\to S_{\nu_{k}}$.
Denote the probability flow associated with $e_{k}$ by $\rho_{k}$, 
i.e., $\rho_{k}=\rho_{\nu_{k}\mu_{k}}$.
Then, from (\ref{eq28}), we get
\begin{align}
\rho_{k}&=q_{\nu_{k}\mu_{k}}p_{\mu_{k}}-q_{\mu_{k}\nu_{k}}p_{\nu_{k}} \nonumber\\
&=\sigma(\nu_{k},\mu_{k})^{T}p.
\label{eq6-6}
\end{align}
The transition matrix $Q'$ corresponding to the reduced loop-free
transition diagram is thus
\begin{equation}
Q'=Q+\sum^{l}_{k=1}\delta(\nu_{k},\mu_{k})\sigma(\nu_{k},\mu_{k})^{T}.
\label{eq33}
\end{equation}
Now we define a vector $\xi(\nu_{k},\mu_{k})$ satisfying
\begin{align}
&Q'\xi(\nu_{k},\mu_{k})=\delta(\nu_{k},\mu_{k}), \quad k=1,2,\cdots,l. \nonumber \\
&\xi(\nu_{k},\mu_{k})_{0}=0.
\label{eq34}
\end{align}
We call $\xi(\nu_{k},\mu_{k})$ the \textit{modifier}
of $L_{k}$ corresponding to the edge $S_{\mu_{k}}\to S_{\nu_{k}}$,
which is a generalization of $\xi$ introduced in the preceding section
satisfying (\ref{eq39eq}) or (\ref{eq44eq}).
Since $Q'$ corresponds to a loop-free transition matirx,
its stationary probability distribution $p'$ is given by
\begin{equation}
p'_{i}=\bar{r}_{i0}p_{0}, \quad i=1,2,\cdots,N.
\label{eq35}
\end{equation}
where $\bar{r}_{i0}$ denotes the transition ratio from
$S_{0}$ to $S_{i}$ taken along a unique path connecting
$S_{0}$ to $S_{i}$ in the reduced loop-free transition diagram.
From (\ref{eq33}) and (\ref{eq34}), we get
\[
Q'\bigl(I-\sum^{l}_{k=1}\xi(\nu_{k},\mu_{k})\sigma(\nu_{k},\mu_{k})^{T}\bigr)=Q.
\]
From $Qp=0$,
\[ 
Q'\bigl(I-\sum^{l}_{k=1}\xi(\nu_{k},\mu_{k})\sigma(\nu_{k},\mu_{k})^{T}\bigr)p=0.
\]
From $Q'p'=0$ and the uniqueness of the stationary distribution,
we have
\[
\bigl(I-\sum^{l}_{k=1}\xi(\nu_{k},\mu_{k})\sigma(\nu_{k},\mu_{k})^{T}\bigr)p=p'.
\]
Taking (\ref{eq28}) into account, we can now 
write the stationary probability distribution explicitly as
\begin{equation}
p=p'+\sum^{l}_{k=1}\rho_{k}\xi(\nu_{k},\mu_{k}),
\label{eq36}
\end{equation}
where $p'$ denotes the stationary probability distribution of the reduced
loop-free transition diagram given by (\ref{eq35}).
This generalizes (\ref{eq40eq}).
Equation (\ref{eq36}) clearly shows the importance
of the probability flows reflecting the loop structure of
the transition diagram.
If the probability flows vanish for each loop,
the stationary distribution is identical to that of reduced
loop-free diagram given in (\ref{eq17}).
The modifier $\xi(\nu_{k},\mu_{k})$ represents the dependence of
the stationary distribution on the probability flow.

\begin{figure}[h]
  \begin{center}
    \includegraphics[width=0.3\textwidth]{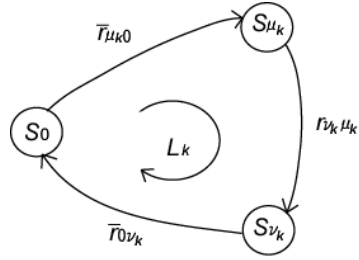}
  \end{center}
  \caption{Loop Elimination by Removal of an Edge.}
  \label{fig:12}
\end{figure}

It remains to compute the probability flow $\rho_{1},\rho_{2},\cdots,\rho_{l}$.
Premultiplication of (\ref{eq36}) by $\sigma(\nu_{k},\mu_{k})^{T}$
yields
\begin{equation}
\rho_{k}-\sum^{l}_{m=1}\sigma(\nu_{k},\mu_{k})^{T}\xi(\nu_{m},\mu_{m})\rho_{m}=\sigma(\nu_{k},\mu_{k})^{T}p'.
\label{eq37}
\end{equation}
Now, from the definition of $\sigma(\nu_{k},\mu_{k})$ (\ref{eq29}),
we have
\begin{align*}
\sigma(\nu_{k},\mu_{k})^{T}p'&=q_{\nu_{k}\mu_{k}}p'_{\mu_{k}}-q_{\mu_{k}\nu_{k}}p'_{\nu_{k}}\\
&=q_{\mu_{k}\nu_{k}}(-p'_{\nu_{k}}+r_{\nu_{k}\mu_{k}}p'_{\mu_{k}})\\
&=-q_{\mu_{k}\nu_{k}}\bar{r}_{\nu_{k}0}(1-\bar{r}_{0\nu_{k}}r_{\nu_{k}\mu_{k}}\bar{r}_{\mu_{k}0})p_{0}
\end{align*}
As is shown in Fig.~\ref{fig:12}, the term 
$\bar{r}_{0\nu_{k}}r_{\nu_{k}\mu_{k}}\bar{r}_{\mu_{k}0}$
corresponds to the product of the transition ratios along $L_{k}$
associated with the edge connecting $S_{\nu_{k}}$ and $S_{\mu_{k}}$,
i.e., the Wegscheider product of $L_{k}$.
We denote it by $\eta_{k}$.
Equation (\ref{eq37}) can now be rewritten as
\begin{equation}
\rho_{k}+\sum^{l}_{m=1}\theta_{km}\rho_{m}=\alpha_{k}(1-\eta_{k})p_{0}, 
\quad k=1,2,\cdots,l,
\label{eqb62}
\end{equation}
where $\theta_{km}=-\sigma(\nu_{k},\mu_{k})^{T}\xi(\nu_{m},\mu_{m})$,
$\alpha_{k}=-q_{\mu_{k}\nu_{k}}\bar{r}_{\nu_{k}0}$, and $\eta_{k}$ is the Wegscheider product of $L_{k}$.
Equation (\ref{eqb62}) implies that the probability flows can be
computed by solving only a linear equation of $l$ unknowns,
instead of $(N+1)$ unknowns of SSE (\ref{eq2-3}) or (\ref{eq2-4a}).
We call (\ref{eqb62}) the \textit{flow equation}.

Now, let us summarize the procedure of computing the stationary 
probability distribution.
\begin{step}\upshape
Find the minimum number of edges $e_{k}: S_{\mu_{k}}\to S_{\nu_{k}}$,
$k=1,2,\cdots,l$ such that their elimination makes the transition
diagram loop-free.
\end{step}
\begin{step}\upshape
Compute the stationary distribution $p'$ via 
(\ref{eq35}) for the reduced loop-free diagram.
\end{step}
\begin{step}\upshape
For each eliminated edge $e_{k}$, associate a loop $L_{k}$,
which is recovered with the edge.
\end{step}
\begin{step}\upshape
\label{step4}
For $e_{k}$, compute the modifier
$\xi(\nu_{k},\mu_{k})$ that satisfies (\ref{eq34}).
\end{step}
\begin{step}\upshape
Solve the flow equation (\ref{eqb62}) to obtain the probability
flows $\rho_{k}$, $k=1,2,\cdots,l$.
\end{step}
\begin{step}\upshape
Compute $p$ from (\ref{eq36}).
\end{step}
\begin{step}\upshape
Compute the exact probability distribution taking the
normalization constraint (\ref{eq2-4}) into account.
\end{step}
The computation of the modifier (Step~\ref{step4}) is discussed
in Appendix.
\begin{ex}\upshape
Consider the transition diagram shown in Fig.~\ref{fig:11}(a) with 
6 states and 3 loops.
To eliminate the loops, two edges, $e_{1}:S_{2}\to S_{3}$
and $e_{2}:S_{4}\to S_{5}$, are removed.
The modified transition diagram is shown in Fig.~\ref{fig:11}(b), 
which is loop-free. We choose  
$L_{1}:S_{0}\to S_{1}\to S_{2}\to S_{3}\to S_{0}$, which is created
by recovering $e_{1}$ in the modified diagram and  
$L_{2}:S_{0}\to S_{3}\to S_{4}\to S_{5}\to S_{0}$ which is created
by recovering $e_{2}$ in the modified diagram as basis loops.
In terms of the notations introduced in this section,
$\mu_{1}=2$, $\nu_{1}=3$, $\mu_{2}=4$, $\nu_{2}=5$,
$\rho_{1}=\rho_{32}$ and $\rho_{2}=\rho_{54}$.
The transition matrices $Q$ and $Q'$ for the original diagram
and the modified diagram are given by
\begin{align*}
&Q=\begin{bmatrix}D_{0}&q_{01}&0&q_{03}&0&q_{05}\\
q_{10}&D_{1}&q_{12}&0&0&0\\
0&q_{21}&D_{2}&\dashbox(15,10){$q_{23}$}&0&0\\
q_{30}&0&\dashbox(15,10){$q_{32}$}&D_{3}&q_{34}&0\\
0&0&0&q_{43}&D_{4}&\dashbox(15,10){$q_{45}$}\\
q_{50}&0&0&0&\dashbox(15,10){$q_{54}$}&D_{5}
\end{bmatrix},\\
&Q'=\begin{bmatrix}D_{0}&q_{01}&0&q_{03}&0&q_{05}\\
q_{10}&D_{1}&q_{12}&0&0&0\\
0&q_{21}&D'_{2}&0&0&0\\
q_{30}&0&0&D'_{3}&q_{34}&0\\
0&0&0&q_{43}&D'_{4}&0\\
q_{50}&0&0&0&0&D'_{5}
\end{bmatrix}
\end{align*}
where $D_{0}=-(q_{10}+q_{30}+q_{50})$, $D_{1}=-(q_{01}+q_{21})$,
$D_{2}=-(q_{12}+q_{32})$, $D_{3}=-(q_{03}+q_{23}+q_{43})$,
$D_{4}=-(q_{34}+q_{54})$, $D_{5}=-(q_{05}+q_{45})$,
$D'_{2}=-q_{12}$, $D'_{3}=-(q_{03}+q_{43})$, $D'_{4}=-q_{34}$,
$D'_{5}=-q_{05}$.
The elements of $Q$ encircled by dashed boxes are eliminated in $Q'$.
The modifiers, $\xi_{1}$ corresponding to $L_{1}$ for
$e_{1}$ and $\xi_{2}$ corresponding to $L_{2}$ for 
$e_{2}$, are given as
\begin{align}
\xi_{1}&=\begin{bmatrix}0\\1/q_{01}\\-r_{21}/q_{01}-1/q_{12}\\
1/q_{03}\\r_{43}/q_{03}\\0\end{bmatrix}, 
\xi_{2}&=\begin{bmatrix}0\\0\\0\\-1/q_{03}\\-r_{43}/q_{03}-1/q_{34}\\1/q_{05}\end{bmatrix}.
\label{eq64eq}
\end{align}
The detail of the above calculation is given in Appendix.

According to (\ref{eqb62}), the flow equation is given by 
\begin{align*}
&(1+\theta_{11})\rho_{1}+\theta_{12}\rho_{2}=\alpha_{1}(1-\eta_{1})\\
&\theta_{21}\rho_{1}+(1+\theta_{22})\rho_{2}=\alpha_{2}(1-\eta_{2})
\end{align*}
where $\theta_{ij}=-\sigma^{T}_{i}\xi_{j}$, $i,j=1,2$, with
\begin{align*}
\sigma^{T}_{1}&=\sigma(3,2)^{T}=
\begin{bmatrix}0&0&q_{32}&-q_{23}&0&0\end{bmatrix}\\
\sigma^{T}_{2}&=\sigma(5,4)^{T}=\begin{bmatrix}0&0&0&q_{54}&-q_{45}&0\end{bmatrix}.
\end{align*}
and
\[
\alpha_{1}=-q_{23}\bar{r}_{30},\quad \alpha_{2}=-q_{45}\bar{r}_{50}.
\]
$\eta_{1}$ and $\eta_{2}$ are Wegscheider products
corresponding to $L_{1}$ and $L_{2}$, respectively, and are
given respectively:
\[
\eta_{1}=r_{10}r_{21}r_{32}r_{03}, \quad \eta_{2}=r_{30}r_{43}r_{54}r_{05}.
\]
\end{ex}

\begin{figure}[h]
  \begin{center}
    \includegraphics[width=0.25\textwidth]{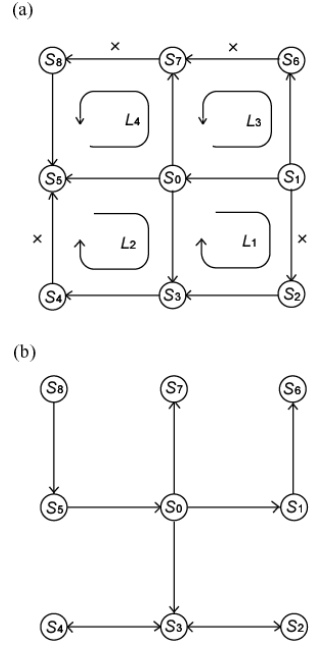}
  \end{center}
  \caption{
An Example of Transition Diagram.
(a) Original Diagram, (b) Reduced Loop-free Diagram.
Transition probabilities are omitted.
}
  \label{fig:13}
\end{figure}

\begin{ex}\upshape
(Enzyme with three effectors)\\
Consider a more complex block transition diagram shown in 
Fig.~\ref{fig:13}(a).
This transition diagram describes an allosteric enzyme with
three different effectors or an operon with three transcription factors.
A description like (\ref{ex3-16}) for this example is very
complicated and hence, is omitted.
Now, we choose four edges $e_{1}: S_{1}\to S_{2}$, $e_{2}: S_{4}\to S_{5}$,
$e_{3}: S_{6}\to S_{7}$ and $e_{4}: S_{7}\to S_{8}$ to make the 
diagram loop-free.
The reduced diagram is shown in Fig.~\ref{fig:13}(b).
A basis set of loops is composed of the following four loops,
$L_{1}: S_{0}\to S_{1}\to S_{2}\to S_{3}\to S_{0}$,  
$L_{2}: S_{0}\to S_{3}\to S_{4}\to S_{5}\to S_{0}$,
$L_{3}: S_{0}\to S_{1}\to S_{6}\to S_{7}\to S_{0}$ and
$L_{4}: S_{0}\to S_{7}\to S_{8}\to S_{5}\to S_{0}$, 
as is shown in Fig.~\ref{fig:13}(a).
The stationary probability distribution corresponding to the reduced
loop-free diagram Fig.~\ref{fig:13}(b) is easily calculated according
to (\ref{eq35}), i.e.,
\begin{align*}
&p'_{1}=r_{10}p_{0},\ p'_{2}=\bar{r}_{20}p_{0}=r_{23}r_{30}p_{0},\ p'_{3}=r_{30}p_{0},\\
&p'_{4}=\bar{r}_{40}p_{0}=r_{43}r_{30}p_{0}, \ p'_{5}=r_{50}p_{0},\\
&p'_{6}=\bar{r}_{60}p_{0}=r_{61}r_{10}p_{0},\ r'_{7}=r_{70}p_{0}, \\ 
&p'_{8}=\bar{r}_{80}p_{0}=r_{85}r_{50}p_{0}.
\end{align*}
The $\sigma $-vectors defined by (\ref{eq29}) in this case are
$\sigma_{1}=\sigma(2,1)$, $\sigma_{2}=\sigma(5,4)$, $\sigma_{3}=\sigma(7,6)$,
$\sigma_{4}=\sigma(8,7)$, i.e.,
\begin{align*}
&\sigma_{1}=\begin{bmatrix}0\\q_{21}\\-q_{12}\\0\\0\\0\\0\\0\\0\end{bmatrix},
\ \sigma_{2}=\begin{bmatrix}0\\0\\0\\0\\q_{54}\\-q_{45}\\0\\0\\0\end{bmatrix},
\ \sigma_{3}=\begin{bmatrix}0\\0\\0\\0\\0\\0\\q_{76}\\-q_{67}\\0\end{bmatrix},
&\sigma_{4}=\begin{bmatrix}0\\0\\0\\0\\0\\0\\0\\q_{87}\\-q_{78}\end{bmatrix}.
\end{align*}
The modifiers $\xi_{i}$, $i=1,2,3,4$ for $L_{1}$, $L_{2}$, $L_{3}$, $L_{4}$
are computed in Appendix; they are
\begin{align}
\xi_{1}&=\begin{bmatrix}0\\-1/q_{01}\\r_{23}/q_{03}+1/q_{32}\\1/q_{03}\\r_{43}/q_{03}\\0\\-r_{61}/q_{01}\\0\\0\end{bmatrix},\ 
\xi_{2}=\begin{bmatrix}0\\0\\r_{23}/q_{03}\\-1/q_{03}\\-r_{43}/q_{03}-1/q_{34}\\1/q_{05}\\0\\0\\r_{85}/q_{05}\end{bmatrix},\ 
\nonumber \\ 
\xi_{3}&=\begin{bmatrix}0\\-1/q_{01}\\0\\0\\0\\0\\-r_{61}/q_{01}-1/q_{16}\\1/q_{07}\\0\end{bmatrix}, \ 
\xi_{4}=\begin{bmatrix}0\\0\\0\\0\\0\\1/q_{05}\\0\\-1/q_{07}\\r_{85}/q_{05}+1/q_{54}\end{bmatrix}
\label{ex8-63}
\end{align} 
The flow equation (\ref{eqb62}) is given by
{\small
\begin{align}
&\begin{bmatrix}
1+\theta_{11}&\theta_{12}&\theta_{13}&0\\
\theta_{21}&1+\theta_{22}&0&\theta_{24}\\
\theta_{31}&0&1+\theta_{33}&\theta_{34}\\
0&\theta_{42}&\theta_{43}&1+\theta_{44}\end{bmatrix}
\begin{bmatrix}\rho_{1}\\\rho_{2}\\\rho_{3}\\\rho_{4}\end{bmatrix} \nonumber \\
&=\begin{bmatrix}\alpha_{1}(1-\eta_{1})\\\alpha_{2}(1-\eta_{2})\\\alpha_{3}(1-\eta_{3})\\\alpha_{4}(1-\eta_{4})\end{bmatrix}
\label{ex8-64}
\end{align}
}
where $\theta_{ij}=-\sigma^{T}_{i}\xi_{j}$, $i=1,2,3,4$, $j=1,2,3,4$,
$\alpha_{1}=-q_{12}\bar{r}_{20}$, $\alpha_{2}=-q_{45}\bar{r}_{50}$,
$\alpha_{3}=-q_{67}\bar{r}_{70}$, $\alpha_{4}=-q_{78}\bar{r}_{80}$, and
$\eta_{i}$, $i=1,2,3,4$, are Wegscheider products for $L_{i}$,
and are given by $\eta_{1}=r_{03}r_{32}r_{21}r_{10}$,
$\eta_{2}=r_{05}r_{54}r_{43}r_{30}$, $\eta_{3}=r_{07}r_{76}r_{61}r_{10}$,
and $\eta_{4}=r_{05}r_{58}r_{87}r_{70}$.
The anti-diagonal elements $\theta_{14}$, $\theta_{23}$,
$\theta_{32}$ and $\theta_{41}$ vanish from the loop structure of
Fig.~\ref{fig:13}.

\end{ex}

\section{Specification of transition probabilities}
\label{sec7}

In order to obtain a useful representation of the activity of 
the biological regulator discussed so far, we must give a way
to connect transition probabilities $q_{ij}$ with more workable
parameters with physical and/or biological meanings.
The state transition caused by binding a BF to a BS and that
caused by releasing a BF from a BS are essentially different
in their nature.
The former transition is usually proportional to the concentration of
the BF to be bound, while the latter does not depend on concentration of
BFs but possibly depends on the occupation pattern
of other BSs.
Therefore, it is natural to assume the following characterization
of transition probabilities:
\begin{equation}
q_{ij}=\left\{%
\begin{array}{ll}
\alpha _{ij}u_{k},& \mbox{if $S_{j}\to S_{i}$ is the binding of $U_{k}$},\\
\alpha_{ij}, & \mbox{if it is a releasing of a BF}
\end{array}\right. 
\label{eqary}
\end{equation}
for each $j$ and $i\in {\bf A}_{j}$, where $\alpha_{ij}$ denotes a 
coefficient and $u_{k}$ the concentration of $U_{k}$.
The transition probabilities used in Example 1 (equation (\ref{eq2-9}))
and these in Example 2 (equation (\ref{eq2-13}))
are special cases of (\ref{eqary}).

From the definition of TR in (\ref{eq14a}), we have
\begin{equation}
r_{ij}=\left\{%
\begin{array}{ll}
\beta_{ij}u_{k}, & \mbox{if $S_{j}\to S_{i}$ is the binding of $U_{k}$},\\
\beta_{ij}u^{-1}_{k}, & \mbox{if it is the releasing of $U_{k}$}.
\end{array}\right.
\end{equation}
where $\beta_{ij}=\alpha_{ij}/\alpha_{ji}$.
We can derive a concrete form for the overall activity of our biological 
regulator based on (\ref{eqary}),
for the case that the Wegscheider condition holds, i.e., the stationary
distribution given by (\ref{eq17}). 

Let $S_{i}$ be a state where $l$ sites are occupied by
$U_{i_{1}}$, $U_{i_{2}}$, $\cdots$, $U_{i_{l}}$
and the remaining $n-l$ sites are empty.
Then, according to (\ref{eq15}) and (\ref{eqary}), 
the probability $p_{i}$ is  
$p_{i}=a_{i}u_{i_{1}}u_{i_{2}}\cdots u_{i_{l}}p_{0}$,
where $a_{i}$ denotes the product of $\beta_{ij}s$ associated with
the binding of $U_{i_{1}}$, $U_{i_{2}}$, $\cdots$, $U_{i_{l}}$.
Note that the binding and unbinding of other BFs cannot take
place because the path connecting $S_{0}$ to $S_{i}$ does not contain
any loop. 
Now, the normalization constraint (\ref{eq2-4}) yields the
overall activity $\gamma$ defined in (\ref{eq2-5}) as 
\begin{equation}
\gamma=\frac{\gamma_{0}+
\Sigma^{N}_{i=1}\gamma_{i}a_{i}u_{i_{1}}u_{i_{2}}
\cdots u_{i_{l}}}
{1+\Sigma^{N}_{i=1}a_{i}u_{i_{1}}u_{i_{2}}
\cdots u_{i_{l}}}. 
\label{eq:b}
\end{equation}
This is a familiar form for describing operon
regulation, which appears frequently in the literature e.g., \cite{mochizuki}
\cite{ozbudak}\cite{segel}.
For the Lac operon of Fig.~\ref{fig:2}, the transition probabilities
given by (\ref{eqary}) implies
$q_{10}=\alpha_{10}u_{1}$, $q_{01}=\alpha_{01}$, $q_{21}=\alpha_{21}u_{2}$,
$q_{12}=\alpha_{12}$, $q_{30}=\alpha_{30}u_{2}$, $q_{03}=\alpha_{03}$,
$q_{40}=\alpha_{40}u_{3}$ and $q_{04}=\alpha_{04}$.
Since only the states bound by $U_{1}$ can initiate transcription,
$\gamma_{0}=\gamma_{3}=\gamma_{4}=0$ in (\ref{eq:b}).
Thus, we have
\begin{equation}
\gamma=\frac{
\gamma_{1}K_{1}u_{1}+
\gamma_{2}K_{1}K_{2}u_{1}u_{2}}
{1+K_{1}u_{1}+K_{1}K_{2}u_{1}u_{2}+K_{3}u_{2}+K_{4}u_{3}}, 
\label{gamfra}
\end{equation}
where $K_{1}=\alpha_{10}/\alpha_{01}$, $K_{2}=\alpha_{21}/\alpha_{12}$,
$K_{3}=\alpha_{30}/\alpha_{03}$ and $K_{4}=\alpha_{40}/\alpha_{04}$. 

The general form (\ref{eq27}), as well as its specification (\ref{eq:b}), 
enables us to derive a variety of complicated formulae representing 
enzymic actions (e.g., \cite{segel}) almost immediately. 
Equation (\ref{eq:b}) also generalizes the classical MWC model \cite{monodj} 
to heterotropic cases.

It is important to notice that the Wegscheider product is
constant and does not depend on concentration of any transcription
factor given the specification (\ref{eqary}).
To see this, notice that 
if the loop contains an edge associated with the binding of $U_{k}$,
it must contain an edge associated with unbinding of $U_{k}$,
because the loop must recover the starting state along the path.
If the binding transition contains a TR with $u_{k}$, then it must contain TR 
associated with unbinding of $U_{k}$ which contains $u^{-1}_{k}$, so that they 
cancel out in the Wegscheider product.
Thus, we have shown that \textit{the Wegscheider product is
always constant} and does not contain the concentrations of BF.

\section{Conclusions}

The metabolic process is controlled by enzymes whose expressions are controlled by genetic regulations. On the other hand, genetic regulation is controlled by the products of metabolism that determine the cell state. In this sense, the genetic and metabolic regulations are closely linked together to form a huge and complex network of intracellular regulations. It has been desirable to establish a common framework for quantitatively describing these regulations in a unified way. 
For that purpose, we used the analogy between genetic and metabolic regulations  shown in Table~\ref{table:1}.  The core of the analogy is that the actual computations of the control action are performed through molecular interactions at the regulatory sites between the sites to be bound and the factors to bind or to dissociate. 

We formulated a finite Markov process describing both the genetic and metabolic regulations. The most salient feature of this Markov process is its reciprocity in transition probability; that is, for each pair of state, if the transition probability from one state to another is positive, the opposite transition probability is also positive. This property reflects the reversibility of the molecular interactions we are dealing with, and it is the source of many interesting properties of the Markov process we have formulated. The parameters that determine the action of regulations is represented as the average rate of the stationary probability distribution of that Markov process, based on the assumption that the dynamics of molecular interactions that compute the control actions are much faster than the reaction they are regulating. This assumption is analogous to the fast equilibrium assumption used in deriving the classical Michaelis-Menten equation \cite{segel}. Actually, our approach can derive a stochastic version of Michaelis-Menten formula  which suggests its legitimacy. Moreover, we can derive a quantitative estimate of how precise the Michaelis-Menten equation is by supplying the variance of the stationary distribution.

We introduced a new notion of probability flow associated with each edge of the transition diagram to represent the loop structure of the transition diagram.  
The state transition diagram is nothing but a representation of the conservation of probability flows at each node.  Based on this observation, we derived many interesting properties of the stationary probability distributions.

We proved a simple result that the probability flows vanish if the graph has no loop. This immediately implies that the detailed balance holds for pairs of  adjacent states. This was generalized to include pairs of states which are not necessarily adjacent to each other. A very simple graphical method of computing the stationary distribution was derived.

We derived a condition which guarantees the detailed balance even if the transition diagram has a loop. This condition is not new; Wegscheider, a German chemist, discovered this condition a century ago. The condition, which we call the Wegsheider condition in this paper, requires that the product of all the transition ratios around a loop be unity. We combined this condition with the probabilistic flow and directly showed that the probability flows vanish under it. 
This result again gives a probabilistic interpretation of this classical result. The detailed balance has been accepted as an obvious fact in literature of chemical kinetics and enzymology, For instance, a famous book of Segel assumed this condition without even mentioning Wegscheider's name \cite{segel}.
We gave a stochastic interpretation of this condition. 

We derived a simple method of computing the stationary probability distribution for general cases with multiple loops.  We showed that if the Wegsheider condition holds for a loop, we can eliminate any of the edges contained in the loop without changing the stationary probability distribution. In other words, if the Wegscheider condition holds for a loop, we can consider a transition diagram with that loop by removing an arbitrary edge within that loop.  The stationary probability distribution of the reduced diagram is identical to that of the original diagram.

We derived a purely graphical method of computing the stationary probability distribution based on the notions of probability flow and modifiers. We derived a simple equation for computing the probability flow. Our method dramatically simplifies the classical King and Altman method \cite{king}. 

Our result suggests that there are two kinds of stationary probability
distributions: the one which satisfies the detailed balance,
and one which does not.
The former class is characterized by zero probability flow, the latter
by non-zero probability flows.
In the latter case, the probability distribution is fixed, but continuous
flows of probability exist in loops, which perhaps reflects
a sort of thermal irreversibility of molecular interactions.
Exploitation of the bio-chemical characterization of the probability flow
is an interesting issue of theoretical biology.

The theoretical framework developed in this paper is based on the 
observation that the intracellular regulations are embedded in a 
homogeneous computational medium of molecular interactions.  
This view is not entirely new, but no serious attempt has been made 
so far to mathematically formulate it, within the best of our knowledge. 
A finite Markov process model proposed in this paper captures some 
essential features of the computational medium and explains the versatility, evolvability and flexibility of the intracellular regulations
and various signal transductions. 
It offers a potential capability to deal with systems in which 
metabolism and gene expressions are linked and integrated together. 
We are now exploiting an analytical tool for investigating 
such systems with greater complexity based on our framework 
presented in this paper.

\section*{Appendix}

Computation of the modifier can be done through the simple recursions
presented in Section~\ref{sec4}.
The modifier corresponding to the loop $L_{k}$ is an
$(N+1)$-dimensional vector $\xi_{k}=\xi(\nu_{k},\mu_{k})$ satisfying
(\ref{eq34}).
For the case that the transition diagram itself is a loop,
we have seen that the recursions formulae (\ref{eq42eq}) and (\ref{eq44plus})
give a vector $\xi$ that satisfies (\ref{eq44eq}),
which corresponds to the special case $\mu_{k}=\nu$, $\nu_{k}=\nu+1$.
The computation of the modifier for a general loop is essentially
reduced to (\ref{eq42eq})(\ref{eq44plus}).
To avoid notational complication,
we assume that $L_{k}$ is the loop $S_{0}\to S_{1}\to S_{2}\to \cdots \to S_{\pi}\to S_{0}$.
Since $e_{k}: S_{\mu_{k}}\to S_{\nu_{k}}$ is in the loop,
we assume $\mu_{k}=\nu$, $\nu_{k}=\nu+1$, $\nu\le \pi-1 $.
Then, the components of the modifier corresponding to the states inside
the loop $L_{k}$ are given by (\ref{eq42eq}) and
(\ref{eq44plus}).
Thus, we have a procedure for computing the components of
the modifier corresponding to the states inside the loop.
The components of $\xi_{k}$ outside the loop
$L_{k}$ are computed as follows: let $S_{j}$ be a state outside
$L_{k}$. Since the reduced diagram is loop-free,
there exists a unique path connecting $S_{j}$ to $S_{0}$.
If the path does not have a common state with $L_{k}$ except
$S_{0}$, let the component of $\xi_{k}$ corresponding to
$S_{j}$ be zero.
If the path has some states in common with $L_{k}$, there is a
state $S_{k_{j}}$ nearest $S_{j}$ in $L_{k}$.
Then, the component of $\xi_{k}$ corresponding to $S_{j}$ is 
given by
\[
(\xi_{k})_{j}=\bar{r}_{jk_{j}}\xi_{k_{j}} \tag{A1}
\]
where $\bar{r}_{jk_{j}}$ is the TR from the state $S_{k_{j}}$ to
$S_{j}$ and $\xi_{k_{j}}$ is the component of the modifier
corresponding to $S_{k_{j}}$ which has already been computed.
The justification of (A1) is based on the remark at the end of
Section~\ref{sec3} (equation (\ref{eq31new})).

\vspace*{5mm}
\noindent
{\bf Example A1.}
We compute $\xi_{1}$ and $\xi_{2}$ in Fig.~\ref{fig:11}.
For $L_{1}$, the forward recursion corresponding to
(\ref{eq42eq}) is given by
\begin{align*}
&\xi_{10}=0 \\
&\xi_{11}=-\frac{1}{q_{01}} \\
&\xi_{12}=r_{21}\xi_{11}-\frac{1}{q_{12}}=-r_{21}\frac{1}{q_{01}}-\frac{1}{q_{12}}
\end{align*}
and the backward recursion corresponding to (\ref{eq44plus})
is given by
\[ \xi_{13}=\frac{1}{q_{03}}. \]
From the graph of Fig.~\ref{fig:11}(b), the path connecting
$S_{4}$ to $S_{0}$ meets $L_{1}$ at $S_{3}$,
while the path connecting $S_{5}$ to $S_{0}$ directly reaches $S_{0}$.
Hence, we have
\[
\xi_{14}=r_{43}\xi_{13}=\frac{r_{43}}{q_{03}},\quad \xi_{15}=0. 
\]
For $L_{2}$, the forward recursion is given by
\begin{align*}
&\xi_{20}=0 \\
&\xi_{23}=-\frac{1}{q_{03}}\\
&\xi_{24}=r_{43}\xi_{23}-\frac{1}{q_{34}}=-\frac{r_{43}}{q_{03}}-\frac{1}{q_{34}}
\end{align*}
and the backward recursion is given by
\[ \xi_{25}=\frac{1}{q_{05}}. \]
For the states outside $L_{2}$, $S_{2}$ and $S_{1}$ are directly
connected to $S_{0}$ without meeting $L_{2}$.
Hence, $\xi_{22}=\xi_{21}=0$. 
Thus, (\ref{eq64eq}) has been confirmed.

\vspace*{5mm}
\noindent
{\bf Example A2.}
We compute $\xi_{1}$, $\xi_{2}$, $\xi_{3}$ and $\xi_{4}$
corresponding to the loops $L_{1}$, $L_{2}$, $L_{3}$ and $L_{4}$ of Example 8.

For $L_{1}$, 
the forward recursion $(S_{0}\to S_{1})$ gives 
\[
\xi_{11}=-1/q_{01},
\] 
and the backward recursion $(S_{0}\to S_{3}\to S_{2})$
gives  
\begin{eqnarray*}
\xi_{13}&=&1/q_{03},\\
\xi_{12}&=&r_{23}\xi_{13}+1/q_{32}.
\end{eqnarray*}
The probabilities of the states outside $L_{1}$ are
given by  
\begin{eqnarray*}
\xi_{14}&=&r_{43}\xi_{13},\\
\xi_{1i}&=&\bar{r}_{i0}\xi_{10}=0,\, i=5,6,7,8.
\end{eqnarray*} 

For $L_{2}$, the 
forward recursion $(S_{0}\to S_{3}\to S_{4})$
gives 
\begin{eqnarray*}
\xi_{23}=-1/q_{03},\\ 
\xi_{24}=r_{43}\xi_{23}-1/q_{34},
\end{eqnarray*}
and the backward recursion $(S_{0}\to S_{5})$
gives 
\[\xi_{25}=1/q_{05}.\] 
The probabilities outside $L_{2}$ 
are given by 
\begin{eqnarray*}
\xi_{22}&=&r_{23}\xi_{23}, \\
\xi_{28}&=&r_{85}\xi_{25}, \\
\xi_{2i}&=&\bar{r}_{i0}\xi_{20}=0,\, i=1,6,7.
\end{eqnarray*}

For $L_{3}$, the forward recursion 
$(S_{0}\to S_{1}\to S_{6})$ gives
\begin{eqnarray*}
\xi_{31}&=&-1/q_{01},\\
\xi_{32}&=&r_{41}\xi_{31}-1/q_{14}, 
\end{eqnarray*}
and the
backward recursion $(S_{0}\to S_{7})$
gives 
\[\xi_{37}=1/q_{07}\]. The probabilities outside $L_{3}$ 
is given by 
\[\xi_{3i}=\bar{r}_{i0}\xi_{20}=0,\, i=2,3,4,5,8\].

For $L_{4}$, the 
forward recursion $(S_{0}\to S_{7})$ 
gives \[\xi_{47}=-1/q_{07}\], and the 
backward recursion $(S_{0}\to S_{5}\to S_{8})$
gives 
\begin{eqnarray*}
\xi_{45}&=&1/q_{05},\\
\xi_{48}&=&r_{85}\xi_{45}+1/q_{54}.
\end{eqnarray*}
The probabilities outside $L_{4}$
is given by \[\xi_{4i}=\bar{r}_{i0}\xi_{40}=0, \, i=1,2,3,4,6.\]

The above scheme verifies (\ref{ex8-63}).

\end{document}